\documentclass{aa}
\usepackage[utf8]{inputenc}
\usepackage[varg]{txfonts}
\usepackage{physics}
\usepackage{gensymb}
\usepackage{nameref}
\usepackage{listings}
\usepackage{balance}
\usepackage{multicol}
\usepackage{multirow}
\usepackage{breqn}
\usepackage{lipsum}
\begin{document}
\title{A kinematically constrained kick distribution\\for isolated neutron stars}
%\titlerunning{}

\author{P. Disberg\inst{1,2,3}\thanks{e-mail: \href{mailto:paul.disberg@monash.edu}{paul.disberg@monash.edu}}
 \and N. Gaspari\inst{1}
 \and A. J. Levan\inst{1,4}
 }

\institute{Department of Astrophysics/IMAPP, Radboud University, P.O. Box 9010, 6500 GL Nijmegen, The Netherlands
\and
School of Physics and Astronomy, Monash University, Clayton, Victoria 3800, Australia
\and
The ARC Center of Excellence for Gravitational Wave Discovery---OzGrav, Australia
\and
Department of Physics, University of Warwick, Coventry CV4 7AL, UK
}

\date{\today}

\abstract
{The magnitudes of the velocity kicks that neutron stars (NSs) obtain at their formation have long been a topic of discussion, with the latest studies analysing the velocities of young pulsars and favouring a bimodal kick distribution.}
{In previous work, a novel method was proposed to determine kicks based on the eccentricity of Galactic trajectories, which is also applicable to older objects. We applied this method to the isolated pulsars with known parallax---both young and old---in order to kinematically constrain the NS natal kick distribution and investigate its proposed bimodality. Since this method is applicable to older pulsars, we effectively increase the sample size with ${\sim}50\%$ compared to the pulsars younger than $10$ Myr.}
{We assumed the velocity vectors of the pulsars to be distributed isotropically in the local standard of rest frame, and for each pulsar we sampled $100$ velocities taking into account this assumption. These velocity vectors were used to trace back the trajectories of the NSs through the Galaxy and estimate their eccentricity. Then, we simulated kicked objects in order to evaluate the relationship between kick magnitude and Galactic eccentricity, which was used to infer the kicks corresponding to the estimated eccentricities.}
{The resulting kick distributions indeed show a bimodal structure for young pulsars and our fits resemble the ones from literature well. However, for older pulsars the bimodality vanishes and instead we find a log-normal kick distribution peaking at ${\sim}200$ km/s and a median of ${\sim}400$ km/s (for velocities below $1000$ km/s). We also compare our methods to literature that suggests natal kicks are significantly higher and follow a Maxwellian with $\sigma=265$ km/s. We cannot reproduce these results using their sample and distance estimates, and instead find kicks that are consistent with our proposed distribution.}{We conclude that our kinematically constrained kick distribution is well-described by a log-normal distribution with $\mu=6.38$ and $\sigma=1.01$, normalised between $0$ and $1000$ km/s. This analysis reveals no evidence for bimodality in the larger sample, and we suggest that the bimodality found by existing literature may be caused by their relatively small sample size.}
\maketitle

\section{Introduction}
\label{sec1}
Neutron stars (NSs) that are observed as isolated pulsars (i.e.\ pulsars without a binary companion) are known to have relatively high velocities compared to their progenitor stars \citep[e.g.][]{Gunn_1970,Lyne_1994,Lorimer_1997,Cordes_1998}, because of which they are thought to have received a (natal) kick at their formation \citep[e.g.][]{Dewey_1987,Bailes_1989,VanParadijs_1995,Iben_1996,Fryer_1997,VandenHeuvel_1997}. These natal kicks are most often associated with anisotropic supernova (SN) explosions \citep[e.g.][]{Janka_1994,Burrows_1995,Burrows_1996,Janka_2017,Gessner_2018,Muller_2019}, although jets launched by the newly-formed NS could also potentially play a significant role \citep[e.g.][]{Soker_2010,Bear_2024,Soker_2024,Wang_2024}. Moreover, some studies argue that a \textquotedblleft rocket\textquotedblright\ mechanism, caused by anisotropic emission of electromagnetic radiation due to the spin-down of the NS, could provide (minor) contributions to the observed velocities \citep[e.g.][]{Harrison_1975,Lai_2001,Agalianou_2023,Igoshev_2023,Hirai_2024}.\\
\indent Core-collapse SNe can not only form NSs but also stellar-mass black holes, which are therefore potential recipients of natal kicks as well \citep[e.g.][]{Brandt_1995,Nelemans_1999,Jonker_2004,Repetto_2012,Mandel_2016,Andrews_2022,Banagiri_2023,Vigna_2024}. In addition, when kicked objects are part of a binary, the system also experiences a Blaauw kick through recoil due to mass-loss \citep{Blaauw_1961,Vandenheuvel_2000}. This means that the total (systemic) kick imparted on a binary consists of a natal kick and a Blaauw kick for each component that has experienced a SN \citep[e.g.][]{Hills_1983,Tauris_2017,Andrews_2019}. Determining the magnitude of NS kicks is therefore not only important for studying the Galactic distribution of NSs \citep[e.g.][]{Prokhorov_1994,Kiel_2009,Ofek_2009,Sartore_2010,Sweeney_2022,Song_2024} but also for the rates and locations of the mergers of compact binaries and the resulting transient signals \citep[e.g.][]{Fryer_1999,Voss_2003,Church_2011,Abbott_2017,Vigna_2018,Zevin_2020,Iorio_2023,Gaspari_2024,Gaspari_2024b,Wagg_2025}, as well as important for understanding the SN mechanism \citep[e.g.][]{Nagakura_2019,Muller_2020,Coleman_2022,Burrows_2024,Kondratyev_2024,Janka_2024} and therefore also the mass distributions of compact objects \citep[e.g.][]{Disberg_2023,Schneider_2023,Laplace_2025}.\\
\indent In an attempt to determine the natal kicks of NSs, several studies have analysed the velocities of (young) pulsars \citep[e.g.][see also Appendix \ref{App.A}]{Arzoumanian_2002,Brisken_2003,Hobbs_2005,Faucher_2006,Verbunt_2017,Igoshev_2020,Disberg_2025}, often by formulating statistical estimates of the (unknown) radial component of the velocity vector relative to the observer. For example, \citet{Hobbs_2005} employed a deconvolution technique \citep[similar to the \lstinline{CLEAN} algorithm of][]{Hogbom_1974} to infer the full 3D velocity from the observed 1D and 2D pulsar velocities, and argued that the velocity distribution of pulsars younger than $3$ Myr approximates the natal kick distribution and is well-described by a Maxwellian distribution with $\sigma=265$ km/s. However, \citet{Verbunt_Cator_2017} find that (1) there are systematic errors because of which the distance estimates used by \citet{Hobbs_2005} that are based on dispersion measures \citep{Cordes_1998,Yao_2017} are likely inaccurate \citep[as confirmed by][]{Deller_2019}, and (2) it is difficult to explain the observed transverse 2D velocities of pulsars using the kick distribution of \citet{Hobbs_2005}.\\
\indent Because of these concerns, \citet{Verbunt_2017} developed a formalism to compare the observed properties of pulsars directly to potential kick distributions, and applied this method to pulsars with distances estimated through parallax instead of dispersion measure. After \citet{Deller_2019} significantly expanded the sample of pulsars with parallax estimates, \citet{Igoshev_2020} applied the method of \citet{Verbunt_2017} to the complete pulsar sample. Both \citet{Verbunt_2017} and \citet{Igoshev_2020} find that a distribution consisting of two Maxwellians provides a better fit to the data than a single Maxwellian (for details see Appendix \ref{App.A}). However, since they compare the proposed distributions directly to observations they stress that this does not necessarily mean that the kick distribution is truly bimodal, it could also mean that the kick distribution is simply wider than a single Maxwellian. After all, for any non-Maxwellian underlying kick distribution adding Maxwellians improves the accuracy of the fit.\\
\indent There are, however, several reasons why the observed NSs could indeed follow a bimodal kick distribution, the first of which is a dichotomy in the formation mechanism. For example, electron-capture supernovae (ECSNe) \citep{Miyaji_1980,Nomoto_1987,Takashi_2013,Doherty_2017} from single stars such as super-AGB stars can account for 2 to 20\% of all SNe \citep{Poelarends_2008,Doherty_2015} and are thought to result in smaller natal kicks compared to the canonical core-collapse SNe \citep[][and references therein]{Vandenheuvel_2010}. Small natal kicks could also be the result of interactions in binary stars, which are relevant since the population of isolated NS contains NSs that are born in binaries and have received a kick large enough to unbind the system \citep[although binary disruptions due to mass loss or natal kicks result in different velocities for the escaped pulsar, see e.g.][]{Beniamini_2024}. Mass transfer in a binary can strip the NS progenitor and lead to stripped and ultra-stripped SNe \citep{Podsiadlowski_2004,Tauris_2013,Tauris_2015}, which in turn result in small natal kicks. However, it has been shown that these smaller kicks would not be sufficient to unbind the binary, and hence that isolated NSs born in binaries can only represent the upper tail of the kick distribution \citep{Kuranov_2009}. More recently, SN simulations suggest that there are likely multiple ranges of zero-age main-sequence masses leading to NS formation \citep[e.g.][]{Ertl_2016,Muller_2016,Kresse_2021}. In particular, \citet{Burrows_2024} find two classes of progenitors: low mass and low compactness that lead to kicks of ${\sim}100{-}200$ km/s and high mass and high compactness that lead to kicks of ${\sim}300{-}1000$ km/s. These two classes could hypothetically correspond to the two Maxwellians found by \citet{Verbunt_2017} and \citet{Igoshev_2020}.\\
\indent The methods of \citet{Hobbs_2005}, \citet{Verbunt_2017}, and \citet{Igoshev_2020} constrain natal kicks by analysing the velocities of young pulsars. This motivation of this choice is that a sample of old pulsars would be biased toward low kick in at least two ways: (1) the Galactic trajectories become more eccentric as a result of the kicks, because of which the objects are more likely to be observed near their Galactic apocentre where they have reduced speeds relative to their initial velocities \citep{Hansen_1997,Disberg_2024a}, and (2) NSs that receive high kicks migrate outwards more quickly and therefore become less likely to be observed as they age \citep{Cordes_1986,Lyne_1994}. In order to eliminate the first of these biases, \citet{Disberg_2024b} proposed a method of inferring kicks (i.e.\ the systemic kicks of binary NSs) based on the shape of their Galactic trajectory, which remains close to circular for low kicks and becomes more eccentric for increasing kick magnitude. This method employs the entire Galactic orbit instead of the present-day circular velocity to infer kicks and should therefore also provide relatively accurate estimates for older NSs. The advantages of such an approach are two-fold. Firstly, the use of older pulsars provides a second, entirely independent sample of neutron stars on which the kick distribution can be determined while accounting for the evolution of their velocity as they migrate through the Galactic potential. Secondly, the older pulsars expand the sample with measured parallax, making it less prone to noise due to low-number statistics compared to the sample containing only young sample with measured parallax, while retaining the advantage of parallax measurements over (likely more uncertain) distance estimates through dispersion measures.\\
\indent We applied the method of \citet{Disberg_2024b} to the isolated pulsars from the samples of \citet{Verbunt_2017} and \citet{Igoshev_2020} in order to kinematically constrain the natal kick distribution of NSs and compare it to the results from literature. In Sect.\ \ref{sec2} we discuss the distance estimates formulated by \citet{Verbunt_Cator_2017} and \citet{Verbunt_2017} and analyse the Galactic trajectories of the pulsars. Then, in Sect.\ \ref{sec3}, we describe and expand the method of \citet{Disberg_2024b}. The resulting kick distributions are shown in Sect.\ \ref{sec4} and are compared to estimates from existing literature. Finally, we summarise our conclusions in Sect.\ \ref{sec5}.
\section{Pulsars}
\label{sec2}
\begin{figure}
    \resizebox{\hsize}{!}{\includegraphics{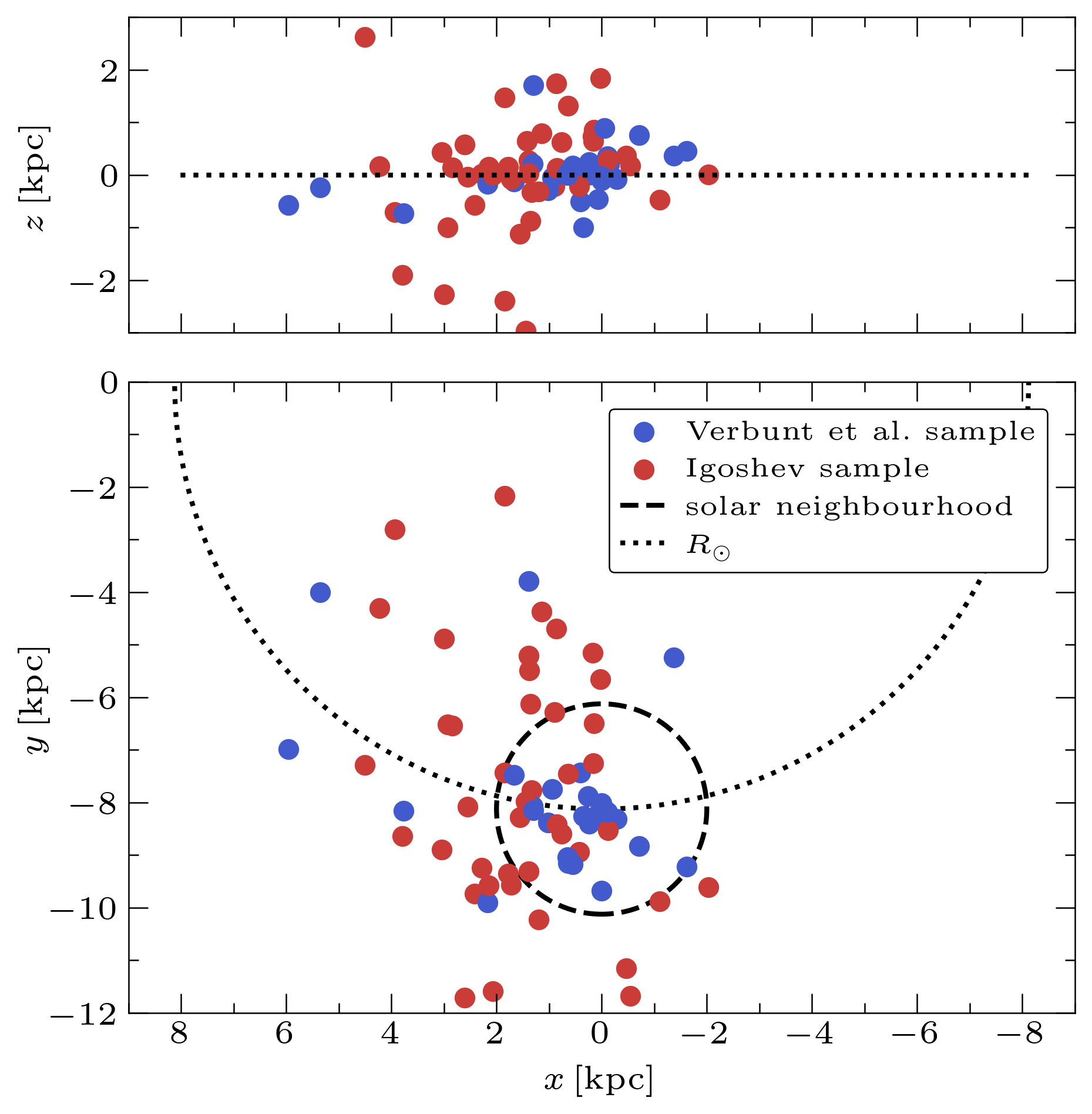}}
    \caption{Median locations of the pulsars in the sample of \citet[][blue dots]{Verbunt_2017} and the expansion made by \citet[][red dots]{Igoshev_2020}, where the distances are determined through their parallaxes (Eq.\ \ref{eq4}). The dashed line shows the solar neighbourhood (i.e.\ $D\leq2$ kpc) whereas the dotted line shows the solar galactocentric radius \citep[$R_{\sun}=8.122$ kpc,][]{Gravity_2018}.} 
    \label{fig1}
\end{figure}
\begin{figure*}
    \sidecaption
    \includegraphics[width=12cm]{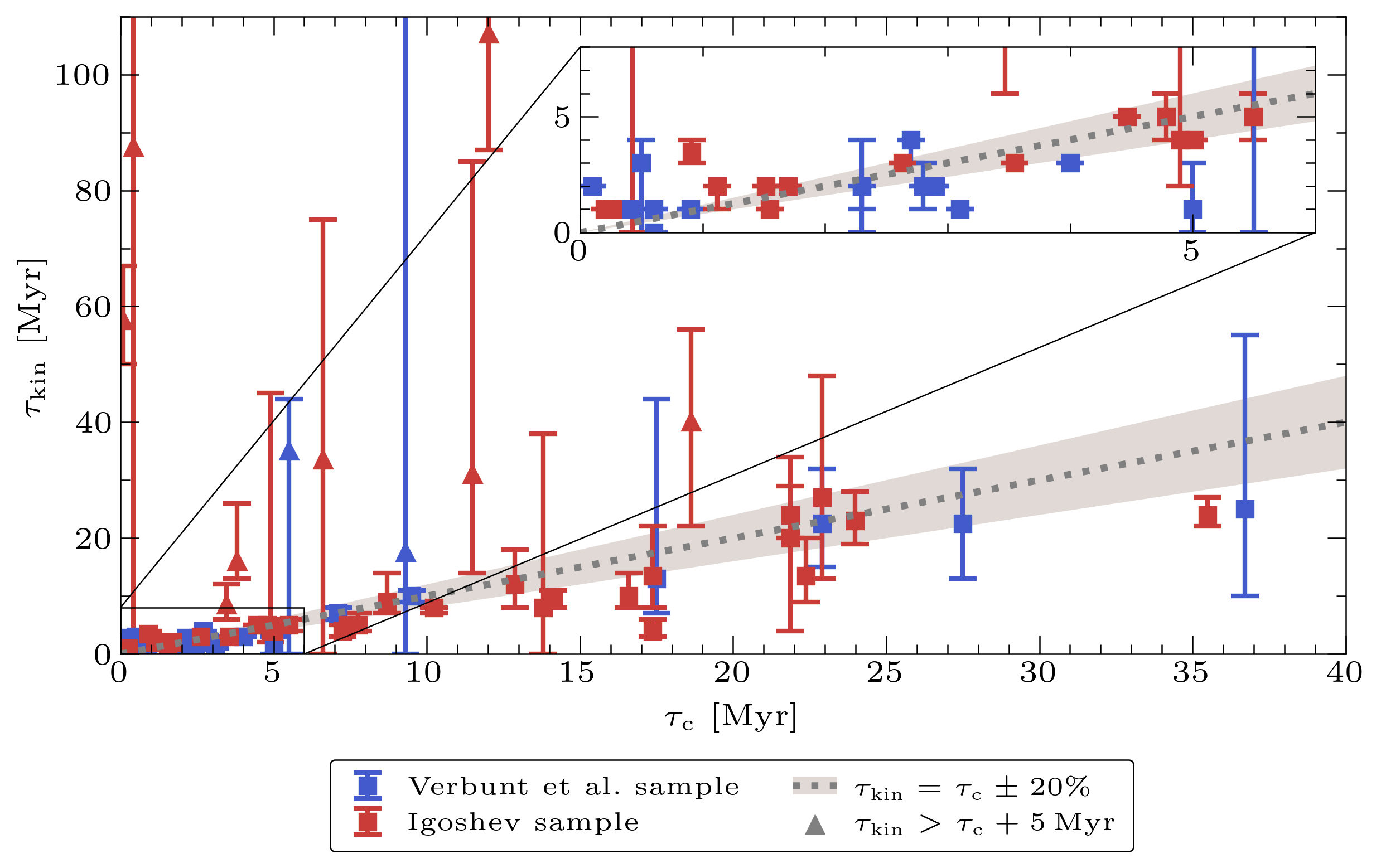}
    \caption{Kinematic ages ($\tau_{\text{kin}}$) versus characteristic spin-down ages ($\tau_{\text{c}}$) for the pulsars in the sample of \citet[][blue]{Verbunt_2017} and the expansion of \citet[][red]{Igoshev_2020} with characteristic ages below $40$ Myr. The kinematic ages were determined through the sets of pulsar trajectories, where the squares are set at the median values and the whiskers show the $68\%$ ranges. The dotted line depicts $\tau_{\text{kin}}=\tau_{\text{c}}$ and the shaded region shows a $20\%$ error on $\tau_{\text{c}}$. If the median kinematic age exceeds the characteristic age with more than $5$ Myr we show it as a triangle instead of a square, because the question arises whether these pulsars affect the accuracy of resulting kick distributions.}   
    \label{fig2}
\end{figure*}
In order to estimate the natal kicks of NSs, we first employed the formalism of \citet[][as used by \citeauthor{Verbunt_2017} \citeyear{Verbunt_2017} and \citeauthor{Igoshev_2020} \citeyear{Igoshev_2020}]{Verbunt_Cator_2017} to determine pulsar distances through their observed parallaxes (Sect.\ \ref{sec2.1}). Then, we estimated the 3D velocity vectors of these pulsars and computed their Galactic trajectories, which allowed us to investigate their kinematic ages as well as their Galactic eccentricities (Sect.\ \ref{sec2.2}).
\subsection{Distances}
\label{sec2.1}
Because of systematic uncertainties in distance estimates through dispersion measures, \citet{Verbunt_2017} limited their analysis to pulsars with known parallax. In order to infer the distance of a pulsar based on its parallax, \citet{Verbunt_Cator_2017} defined the conditional probability of observing a parallax equal to $\omega_{\text{obs}}$ when the true distance equals $D$ as:
\begin{equation}
    \label{eq1}
    P(\omega_{\text{obs}}\text{\hspace{.4mm}}|\text{\hspace{.4mm}}D)\propto\exp\left(-\dfrac{\left(\left[D/\text{kpc}\right]^{-1}-\left[\omega_{\text{obs}}/\text{mas}\right]\right)^2}{2\left[\sigma_\omega/\text{mas}\right]^2}\right),
\end{equation}
where $\sigma_{\omega}$ is the measurement error for the parallax. Then, they defined the prior probability of observing a pulsar with a distance $D$. If a pulsar is observed at a certain sky location, the distance probability is not uniformly distributed but follows the density of the Galactic pulsar population. For this reason, \citet{Verbunt_Cator_2017} employed the prior distance function from \citet{Verbiest_2012}, which equals \citep[in the notation of][]{Igoshev_2016}:
\begin{equation}
    \label{eq2}
    P(D)\propto D^2R^{1.9}\exp\left(-\dfrac{|z|}{0.33\ \text{kpc}}-\dfrac{R}{1.70\ \text{kpc}}\right) ,
\end{equation}
where $z$ and $R$ are the galactocentric height and cylindrical radius of the pulsar, respectively. These values are related to the distance and the observed sky locations (i.e.\ the galactic coordinates $l$ and $b$) through: 
\begin{equation}
    \label{eq3}
    z=D\sin b;\ \text{ and }\ R=\sqrt{R_{\sun}^2+\left(D\cos b\right)^2-2R_{\sun}D\cos b\cos l},
\end{equation}
with $R_{\sun}$ the galactocentric cylindrical radius of the Sun, which we take to be equal to $8.122$ kpc \citep{Gravity_2018}. Since $P(\omega_{\text{obs}}\text{\hspace{.4mm}}|\text{\hspace{.4mm}}D)$ is relatively well-constrained due to the accuracy of parallax measurements, the values of the parameters in Eq.\ \ref{eq2} do not affect the inferred distances significantly \citep[as shown by][]{Igoshev_2020}. The distance to a pulsar is inferred through \citep[cf.\ Eq.\ 5 of][]{Verbunt_Cator_2017}:
\begin{equation}
    \label{eq4}
    P(D\text{\hspace{.4mm}}|\text{\hspace{.4mm}}\omega_{\text{obs}})\propto P(D)\,P(\omega_{\text{obs}}\text{\hspace{.4mm}}|\text{\hspace{.4mm}}D)
\end{equation}
which is a function of $D$ depending on the observed parallax ($\omega_{\text{obs}}$) and the sky coordinates $l$ and $b$ (through Eq.\ \ref{eq3}).\\
\indent We considered the \citet{Verbunt_2017} pulsar sample \citep[based on data from][taken from the ATNF catalogue\footnote{\href{https://www.atnf.csiro.au/research/pulsar/psrcat}{https://www.atnf.csiro.au/research/pulsar/psrcat}.} of \citeauthor{Manchester_2005} \citeyear{Manchester_2005}]{Chatterjee_2001,Brisken_2002,Brisken_2003b,Chatterjee_2004,Chatterjee_2009,Deller_2009,Kirsten_2015} and the \citet{Igoshev_2020} sample \citep[which more than doubles the Verbunt et al.\ sample by adding data from][]{Deller_2019}. This means that the Verbunt et al.\ sample is a subset of the Igoshev sample. We computed the probability distribution of their distances through Eq.\ \ref{eq4}, and for each pulsar we sampled $10^2$ distances from this distribution. These distance estimates were combined with the observed sky locations to determine their galactocentric locations. In Fig.\ \ref{fig1} we show the median locations of the pulsars in both samples (where we only show the pulsars in the Igoshev sample that are not in the Verbunt et al.\ sample). In Appendix \ref{app.B} we show the distributions of the pulsar distances, where ${\sim}80\%$ of the Verbunt et al.\ sample and ${\sim}60\%$ of the (total) Igoshev sample is located in the solar neighbourhood (which we defined as $D\leq2$ kpc). Moreover, the majority of pulsars are located at $x>0$ kpc \citep[cf. Fig.\ 5 of][]{Chrimes_2021}. This is also noted by \citet{Igoshev_2020}, who suggest this might be due to some regions of the Galaxy being unavailable for very long baseline interferometry (VLBI) measurements.
\subsection{Trajectories}
\label{sec2.2}
\begin{figure*}
    \centering
    \includegraphics[width=18cm]{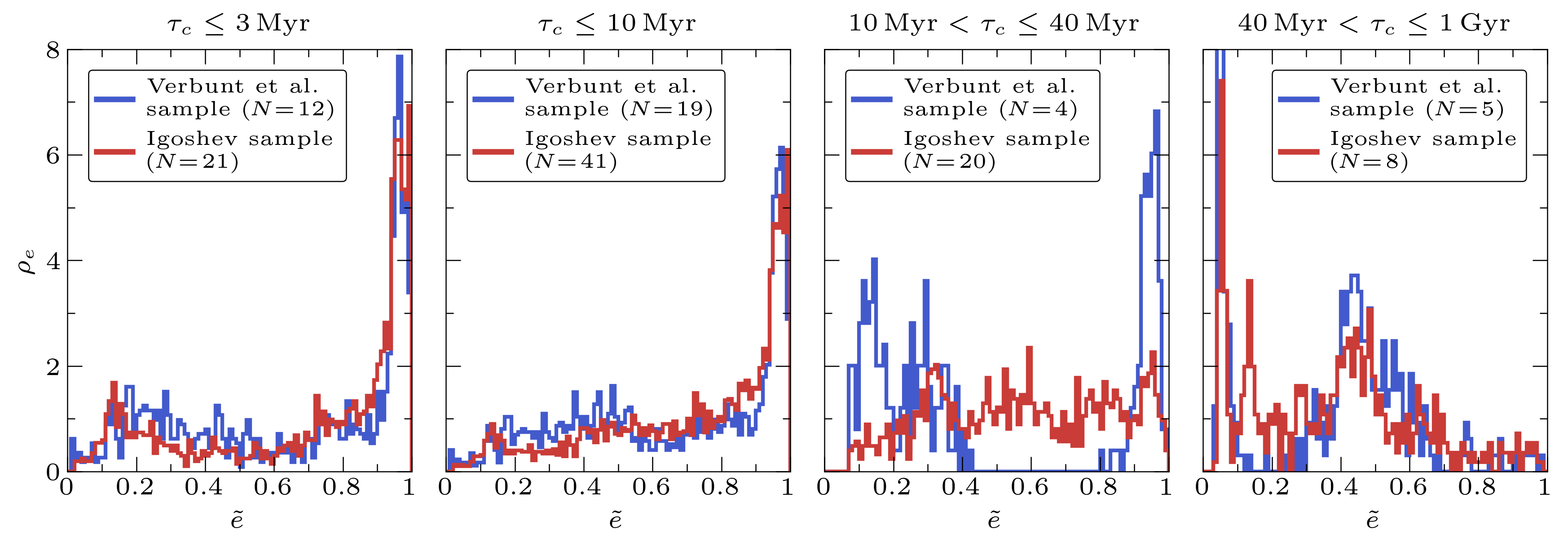}
    \caption{Galactic eccentricities for the pulsars in the samples of \citet[][blue]{Verbunt_2017} and \citet[][red]{Igoshev_2020}, as defined in Eq.\ \ref{eq6}. The distributions are shown in histograms with bins of $0.01$, and divided in four ranges of $\tau_{\text{c}}$ consisting of $N$ pulsars.}
    \label{fig3}
\end{figure*}
Having determined the distances to the pulsars in the Verbunt et al.\ and the Igoshev samples, we turned to the observed proper motions in order to estimate their velocities. That is, the proper motions can be combined with the distances to obtain the 2D transverse velocity vectors (i.e.\ $\vec{v}_{t}$) which are the projection of the full 3D velocity vector on the sky (and corrected for the transverse component of the Sun's velocity). However, the radial component of the velocity vector (i.e.\ $v_r$) cannot be constrained from observations, so in order to estimate the pulsar velocities we follow the method of \citet{Gaspari_2024} in assuming isotropy to determine the probability distribution of $v_r$. They compared two options: isotropy in the galactocentric frame (GC isotropy) and isotropy in the local standard of rest frame (LSR isotropy). However, \citet{Disberg_2024b} showed that LSR isotropy is a more accurate assumption (for any given kick magnitude). For this reason we choose LSR isotropy to determine the pulsar velocity vectors. Following this assumption, the radial component of the velocity vector is estimated through:
\begin{equation}
    \label{eq5}
    v_r=\left|\left|\vec{v}_t-\vec{v}_{\text{LSR},\,t}\right|\right|\hspace{.3mm}\cot\theta+v_{\text{LSR},\,r},
\end{equation}
where $\vec{v}_{\text{LSR},\,t}$ and $v_{\text{LSR},\,r}$ are the transverse and radial components of the LSR velocity, respectively, and $\theta=\arccos u$ with $u$ being uniformly sampled between $-1$ and $1$. For each sampled pulsar location, we took a uniformly sampled value of $u$ and calculated $v_r$ which combined with $\vec{v}_t$ determines the velocity vector.\\
\indent Having obtained $10^2$ locations and velocities for each pulsars, we flipped the estimated velocity vectors and used \lstinline{GALPY}\footnote{\href{http://github.com/jobovy/galpy}{http://github.com/jobovy/galpy}.} \lstinline{v.1.10.0} \citep{Bovy_2015} to trace back the trajectories through the Milky Way using the Galactic potential of \citet{McMillan_2017}, where we---similarly to \citet{Disberg_2024a,Disberg_2024b}---set the circular velocity at $R_{\sun}$ equal to $v_{\phi,\,\sun}=245.6$ km/s \citep{Gravity_2018}. We computed the pulsar trajectories for $1$ Gyr and evaluated them every $1$ Myr.\\
\indent The sets of pulsar trajectories allowed us to estimate the kinematic ages ($\tau_{\text{kin}}$) of the pulsars, which are defined as the time that has passed since their last disc crossing (i.e.\ at $z=0$ kpc). After all, since NSs are formed in SNe of massive stars, which are located in the thin disc at $z\approx0$ kpc, the last disc crossing can provide a lower bound on the age of a NS. The kinematic ages can, then, be compared to the characteristic spin-down ages of the pulsars ($\tau_{\text{c}}$), which are defined through their rotational period and its derivative \citep[$\tau_{\text{c}}=P/(2\dot{P})$, e.g.][]{Shapiro_1983}. Although in some cases it seems to be inaccurate \citep[e.g.][]{Jiang_2013,Zhang_2022}, the characteristic age is usually considered a reasonable estimate of the true pulsar age \citep[as e.g.\ argued by][]{Maoz_2024}. Several studies have compared kinematic and characteristic ages \citep[e.g.][]{Lyne_1982,Cordes_1998,Brisken_2003} and in particular \citet{Noutsos_2013} and \citet{Igoshev_2019} find a general agreement between the two (independent) age estimates. Here, we compared $\tau_{\text{kin}}$ to $\tau_{\text{c}}$ in order to test whether the assumption of LSR isotropy in the distribution of $v_r$ (which influences $\tau_{\text{kin}}$) is consistent with the assumptions behind $\tau_{\text{c}}$. In Fig.\ \ref{fig2} we show the kinematic age distributions----that follow from the sets of pulsar trajectories---versus the characteristic ages. Indeed, the assumption of LSR isotropy results in the majority of the median kinematic age estimates to differ less than ${\sim}20\%$ from the corresponding characteristic ages. There are, however, a few pulsars with $\tau_{\text{kin}}$ exceeding $\tau_{\text{c}}$ substantially, which could either mean that the assumption of LSR isotropy is invalid or that $\tau_{\text{c}}$ is not indicative of the true age \citep[e.g.\ due to post-SN fallback on the NS, as discussed by][]{Igoshev_2016a}. In particular, we considered pulsars where median $\tau_{\text{kin}}$ exceeds $\tau_{\text{c}}$ with more than $5$ Myr to have potentially uncertain ages. Nevertheless, in general we deem the characteristic age a reliable age estimate and consistent with the assumption of LSR isotropy.\\
\indent In order to use the method of \citet{Disberg_2024b} to kinematically constrain the natal kick of the pulsars, we first investigated the shape (i.e.\ eccentricity) of their Galactic trajectories. After all, a larger kick disturbs the initial circular orbits more and therefore results in a more eccentric path through the Galaxy. Since we have established that $\tau_{\text{c}}$ is a reasonable age estimate, we limited our analysis to the pulsar trajectories that do not deviate from the Galactic disc when traced back for a period equal to the characteristic age. That is, we conservatively adopted the constraint that the traced-back trajectories should be positioned within $R\leq20$ kpc and $\text{|}z\text{|}\leq1$ kpc at $t=\tau_{\text{c}}$. Although for a few pulsars this decreases the amount of trajectories significantly (i.e.\ J1321+8323, J1543+0929, J1840+5640, J2046+1540, J2046--0421, J2248--0101, J2346--0609), the median exclusion rate is only $3\%$. We evaluate the cylindrical radii (i.e.\ the Galactic distances projected on the Galactic plane) of the selected trajectories and take their minimum and maximum values ($R_{\min}$ and $R_{\max}$, respectively). The eccentricity of the Galactic orbit is then defined as: 
\begin{equation}
    \label{eq6}
    \tilde{e}=\dfrac{R_{\max}-R_{\min}}{R_{\max}+R_{\min}},
\end{equation}
analogous to a Keplerian eccentricity (although the trajectories are not Keplerian due to the Galactic potential). If an object that starts in a circular Galactic orbits experiences no kick, then $R_{\min}=R_{\max}$ and $\tilde{e}=0$. In contrast, if the object receives an extremely large kick, it will escape the Galaxy meaning $R_{\max}\gg R_{\min}$ and $\tilde{e}\approx1$.\\
\indent In Fig.\ \ref{fig3} we show the resulting eccentricity distributions \citep[cf.\ Fig.\ 9 of][]{Disberg_2024b} for four subsets: the \textquotedblleft young\textquotedblright\ pulsars as defined by \citet[][i.e.\ $\tau_{\text{c}}\leq3$ Myr]{Igoshev_2020} and \citet[][i.e.\ $\tau_{\text{c}}\leq10$ Myr]{Verbunt_2017}, as well as old pulsars (i.e.\ $40$ Myr $\leq\tau_{\text{c}}\leq1$ Gyr) that have obtained (galactocentric) velocities ${\sim}200$ km/s \citep[independent of kick distribution, see Appendix B of][]{Disberg_2024a} and pulsars in between these two categories \citep[i.e.\ $10$ Myr $\leq\tau_{\text{c}}\leq40$ Myr, see Appendix A of][]{Disberg_2024a}. We note that the oldest pulsar in the samples has a characteristic age of $\tau_{\text{c}}=575$ Myr, but we set the range of the oldest bin to $1$ Gyr for consistency with the simulation introduced in Sect.\ \ref{sec3}. The eccentricity distributions shown in Fig.\ \ref{fig3} peak at $\tilde{e}\approx1$ for the young pulsars. For older pulsars this peak starts to shift to lower eccentricities, in fact the oldest pulsars peak at $\tilde{e}\approx0$. The small eccentricities indicate that if older pulsars were observed to have small velocities, then these cannot be explained by the deceleration due to galactic drift \citep[as discussed by][]{Hansen_1997,Disberg_2024a} because this would require eccentric orbits.
This means that the shift is likely due to the fact that high kicks displace NSs to relatively large offsets, making them less likely to be observed after a certain amount of time and therefore introducing a selection effect \citep[as discussed by][]{Cordes_1986,Lyne_1994}. Besides this general shift to lower values of $\tilde{e}$, the eccentricity distributions encode the natal NS kicks (see e.g.\ the peak at $\tilde{e}\approx0.45$ for the oldest pulsars). Lastly, we note that applying the method of \citet{Disberg_2024b} to these pulsars will result in more accurate kick estimates for older pulsars (i.e.\ $\tau_{\text{c}}\gtrsim10$ Myr) compared to methods that only estimate their current velocities \citep[e.g.][]{Hobbs_2005,Verbunt_2017,Igoshev_2020}. At the cost of having to consider selection effects (such as NSs that receive large kicks and escape the Galaxy not being observable after a certain amount of time), this method effectively expands the sample of pulsars with relatively accurate kick estimates, making it more robust against noise caused by low-number statistics.
\section{Method}
\label{sec3}
Having evaluated the eccentricities of the pulsars' trajectories, we employed the method of \citet{Disberg_2024b} to estimate their kicks. In order to do this, we expanded their simulation which determines the relationship between Galactic eccentricity and kick magnitude (Sect.\ \ref{sec3.1}). Then, we discussed how we inferred kicks based on the simulation and the eccentricity estimates (Sect.\ \ref{sec3.2}).
\begin{figure*}
    \centering
    \includegraphics[width=18cm]{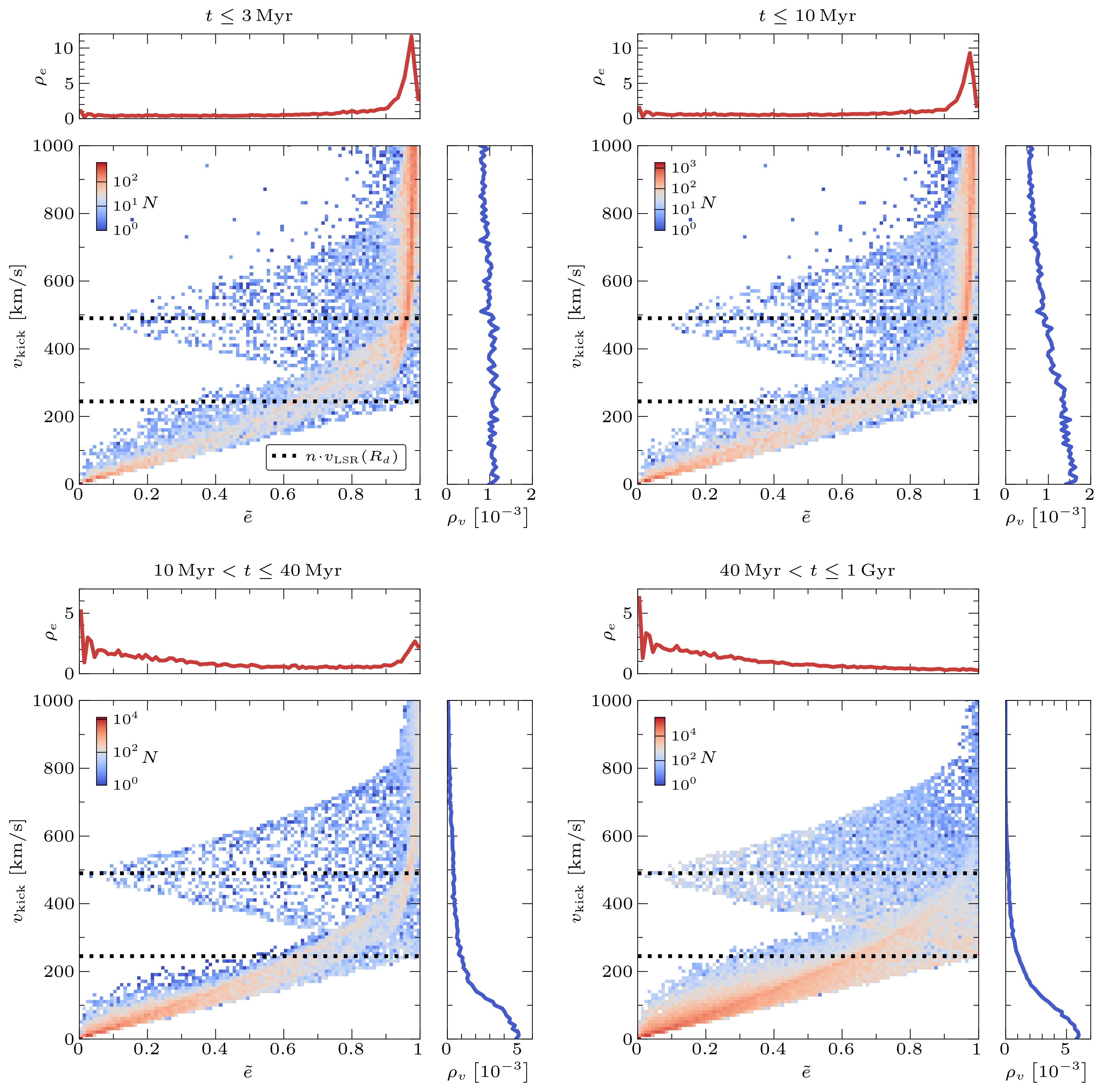}
    \caption{Distributions of $v_{\text{kick}}$ versus $\tilde{e}$ for objects in the solar neighbourhood ($D\leq2$ kpc) as resulting from the simulations, shown in 2D histograms on a logarithmic colour scale with eccentricity bins of $0.01$ and velocity bins of $10$ km/s. We show the eccentricities found in four time-bins, corresponding to the bins from Fig.\ \ref{fig3}. Each distribution also shows the integrated $\tilde{e}$ and $v_{\text{kick}}$ densities ($\rho_e$ and $\rho_v$, respectively). The black dotted lines show one and two times the circular velocity at $R_d=7.04$ kpc (from Eq.\ \ref{eq7}). For the distribution of the total population see Appendix \ref{App.C}.}
    \label{fig4}
\end{figure*}
\subsection{Simulation}
\label{sec3.1}
In order to relate the estimated eccentricities to kicks, we simulated populations of objects receiving kicks with different magnitudes and analysed the eccentricities of the resulting orbits. For each simulation, we seeded $10^3$ objects in a Gaussian annulus at $z=0$ kpc, described by:
\begin{equation}
    \label{eq7}
    P(R)\propto\exp\left(-\dfrac{(R-R_d)^2}{2{\sigma_d}^2}\right).
\end{equation}
\citet{Faucher_2006} first proposed this distribution for pulsars, which was fitted by \citet{Sartore_2010} to the pulsar distribution of \citet{Yusifov_2004} resulting in $R_d=7.04$ kpc and $\sigma_d=1.83$ kpc. We note that a completely different initial distribution (i.e.\ an exponential disc) likely results in similar kick estimates \citep{Disberg_2024b}. Having obtained the initial positions of the objects, we gave them a circular (azimuthal) velocity as determined by the Galactic potential of \citet{McMillan_2017}, and added an isotropically sampled kick velocity with a magnitude $v_{\text{kick}}$. Based on the initial positions and velocities, we computed the Galactic trajectories of the sampled objects for $1$ Gyr \citep[using \lstinline{GALPY} of][]{Bovy_2015}, after which each object was assigned an eccentricity through Eq.\ \ref{eq6}. In order to track the population that is in the solar neighbourhood (i.e.\ $D\leq2$ kpc) at any given time, we initialised the orbit of the Sun \citep[using the results of][see also Eqs.\ 3 and 4 of \citeauthor{Disberg_2024a} \citeyear{Disberg_2024a}]{Reid_2004,Drimmel_2018,Gravity_2018,Bennett_2019} and selected objects that are within $2$ kpc at a given time. After all, it is not that for an individual object its value of $\tilde{e}$ changes over time, but the population of objects that is located in the solar neighbourhood does. We then repeated the simulation for different values of $v_{\text{kick}}$ with intervals of $10$ km/s up to $1000$ km/s \citep[corresponding to the kick ranges found by][]{Burrows_2024}, obtaining time-dependent relationships between $v_{\text{kick}}$ and $\tilde{e}$ for the solar neighbourhood population.\\
\indent In Fig.\ \ref{fig4} we show the resulting $v_{\text{kick}}$ versus $\tilde{e}$ distributions \citep[cf.\ Fig.\ 10 of][]{Disberg_2024b}, for time intervals corresponding to the $\tau_{\text{c}}$ ranges in Fig.\ \ref{fig3}. For all time intervals there is a close relationship between kick and eccentricity for $v_{\text{kick}}\lesssim400$ km/s, meaning our method is most sensitive in this region. For higher kicks the objects pile up just below $1$, since they are likely to escape the Galaxy. After all, while the true Galactic eccentricity equals $1$ for escaping objects, the estimated value of $\tilde{e}$ is limited by the integration time of the simulation. Moreover, \citet{Disberg_2024b} distinguish between objects that follow the rotational direction of the Sun (i.e.\ \textquotedblleft prograde\textquotedblright\ orbits) and objects that travel against the direction of the Solar System (i.e.\ \textquotedblleft retrograde\textquotedblright\ orbits), and these two kinds of orbits can also be found in Fig,\ \ref{fig4}. That is, for $v_{\text{kick}}\lesssim v_{\text{LSR}}(R_d)$ all objects are prograde since the kick is likely smaller than the initial circular velocity of the object. At higher kicks it becomes possible to receive a kick larger than the initial velocity but in opposite direction, resulting in a retrograde orbit. This peaks at $v_{\text{kick}}\approx 2\cdot v_{\text{LSR}}(R_d)$, which can result in a circular (i.e.\ $\tilde{e}=0$) retrograde orbit \citep[for a more elaborate discussion see][]{Disberg_2024b}.\\
\indent Also, Fig.\ \ref{fig4} shows how over time the contribution of high kicks to the solar neighbourhood population decreases. This supports our previous statement on the small eccentricities of Galactic pulsars, namely that they are caused by the selection effect proposed by \citet{Cordes_1986} and \citet{Lyne_1994} by which highly kicked pulsars disappear from the solar neighbourhood over time. This selection effect causes a peak at $\tilde{e}\approx1$, which over time shifts to a peak at $\tilde{e}\approx0$ (for a uniform kick distribution), similarly to the observed pulsar eccentricities shown in Fig.\ \ref{fig3}. Lastly, for older pulsars the $\rho_e$ density displays an oscillation at low $\tilde{e}$, this is not physical but a consequence of the $\tilde{e}$ peak shifting significantly between subsequent kick bins at $v_{\text{kick}}\lesssim30$ km/s.
\subsection{Inference}
\label{sec3.2}
\begin{figure}
    \resizebox{\hsize}{!}{\includegraphics{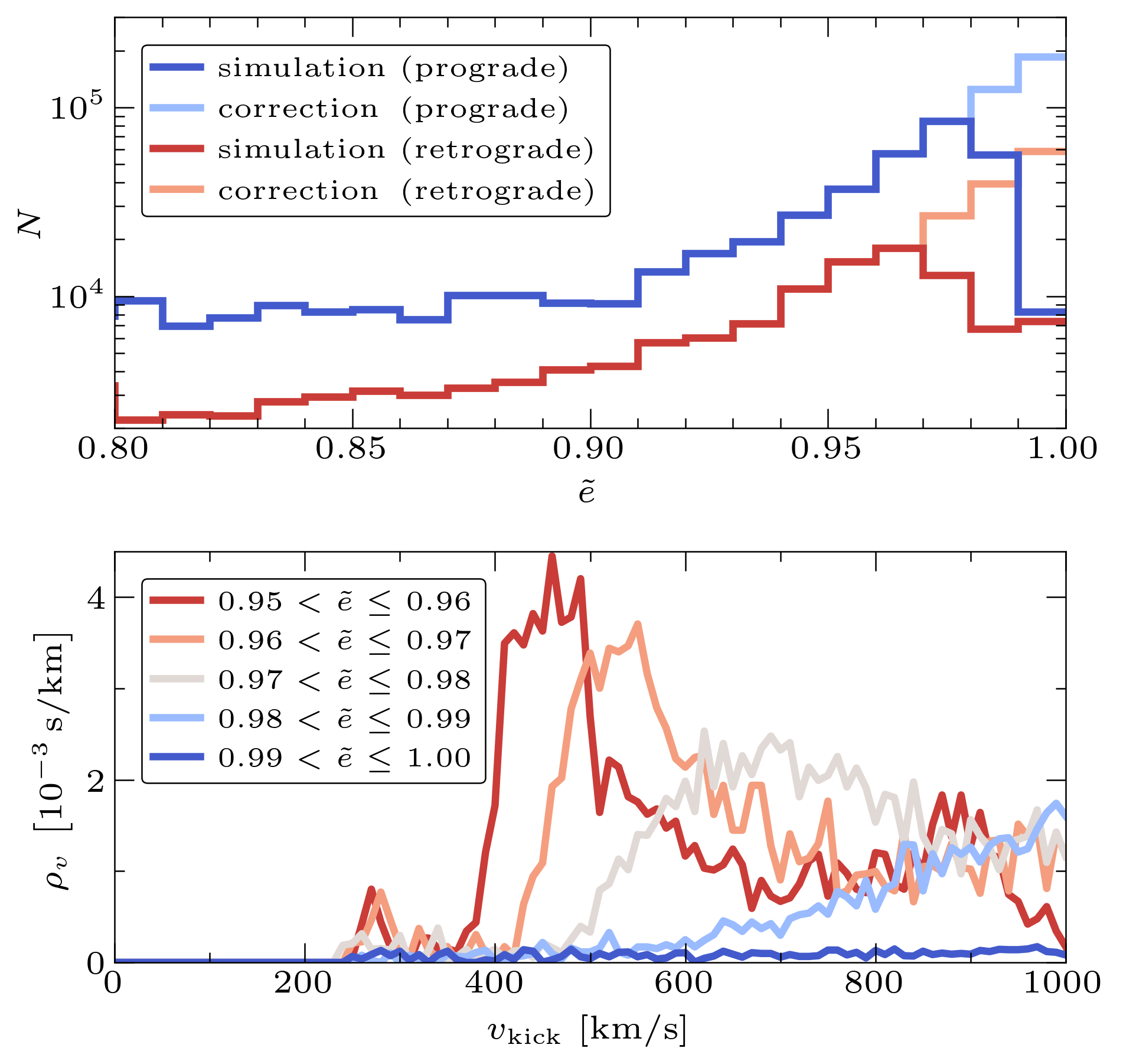}}
    \caption{Correction made to the normalisation of the simulated $v_{\text{kick}}$ versus $\tilde{e}$ distributions of young objects (the figure shows the example of $\tau_{\text{c}}\leq10$ Myr). The top panel shows the amount of objects per eccentricity bin for progrades (blue) and retrogrades (red) orbits, as well as the corrected amounts used for the normalisation (lightblue and lightred, respectively). The bottom panel shows the kick distributions for the five $\tilde{e}$ bins closest to $1$ (indicated by colour), normalised using the corrected values shown above.}
    \label{fig5}
\end{figure}
We have estimated the eccentricities of the observed pulsars (Fig.\ \ref{fig3}) and investigated the relationship between eccentricity and kick magnitude (Fig.\ \ref{fig4}), and we used this to infer the kicks corresponding to the observed eccentricities. That is, the simulations provide insight into the amount of eccentric orbits given a certain kick magnitude but we are interested in inferring the kick magnitude given a certain eccentricity. In order to do this we employed Bayes' theorem, for which the prior distributions of the eccentricities and kicks are needed \citep[see also Appendix B of][]{Disberg_2024b}. First, we divided the simulations in time bins corresponding to the ranges shown in Figs.\ \ref{fig3} and \ref{fig4}, and separated prograde from retrograde orbits (where we note that ${\sim}90\%$ of the observed pulsar trajectories are prograde). For a value of $\tilde{e}$ from the estimated distributions shown in Fig.\ \ref{fig3}, we compare it with simulated results that are in the same time bin as the corresponding $\tau_{\text{c}}$ and orbit the Galaxy in the same rotational direction. Then, the prior probability of observing an eccentricity $\tilde{e}$ is proportional to the amount of simulated orbits with this eccentricity: $P(\tilde{e})\propto N(\tilde{e})$. Likewise, the prior probability of observing an object that received a kick equal to $v_{\text{kick}}$ is given by $P(v_{\text{kick}})\propto N(v_{\text{kick}})$. Also, the probability of observing a value of $\tilde{e}$ given a certain $v_{\text{kick}}$ is the amount of eccentricities in the corresponding bin normalised for objects that have received this kick: $P(\tilde{e}\text{\hspace{.4mm}}|\text{\hspace{.4mm}}v_{\text{kick}})\propto N(\tilde{e}\text{\hspace{.4mm}}|\text{\hspace{.4mm}}v_{\text{kick}})/N(v_{\text{kick}})$. In other words, this conditional probability is proportional to the value of a cell in Fig.\ \ref{fig4} when each row is normalised individually. The defined probabilities can then be inserted into Bayes' theorem, which gives:
\begin{dmath}
    \label{eq8}
    P(v_{\text{kick}}\text{\hspace{.4mm}}|\text{\hspace{.4mm}}\tilde{e})=P(\tilde{e}\text{\hspace{.4mm}}|\text{\hspace{.4mm}}v_{\text{kick}})\dfrac{P(v_{\text{kick}})}{P(\tilde{e})}\propto\dfrac{N(\tilde{e}\text{\hspace{.4mm}}|\text{\hspace{.4mm}}v_{\text{kick}})}{N(\tilde{e})}.
\end{dmath}
That is, the probability that a kick of magnitude $v_{\text{kick}}$ is the cause of an observed (either prograde or retrograde) orbit with an eccentricity $\tilde{e}$ is given by the value of a cell in Fig.\ \ref{fig4} where the columns are normalised individually.\\
\indent However, for the most eccentric orbits the normalisation is not straightforward, because the corresponding kick distribution starts to shift to $v_{\text{kick}}>1000$ km/s and this region is not included in the simulations. For this reason, we normalise these distributions as if the amount of orbits per $\tilde{e}$ bin does not decrease, extrapolating the preceding trend with $d\log_{10}(N)/d\tilde{e}=17$. In the top panel of Fig.\ \ref{fig5} we show this correction in the amounts of objects per eccentricity bin considered for the normalisation. In the bottom panel of the figure we show how the kick distributions per $\tilde{e}$ (normalised using this correction) shift to higher kick magnitudes. Without the correction these renormalised distributions would cause unnecessary noise in the kick estimates.\\
\indent For each individual pulsar there is not one single value of $\tilde{e}$ but a distribution of eccentricities (i.e.\ $D_{\tilde{e}}$) which defines the probability density of $\tilde{e}$ (i.e.\ $P(\tilde{e}\text{\hspace{.4mm}}|\text{\hspace{.4mm}}D_{\tilde{e}})$). The sum of these distributions were shown in Fig.\ \ref{fig3} (for certain ranges of $\tau_{\text{c}}$). The individual kick distributions as constrained through the eccentricity estimates (i.e.\ $P(v_{\text{kick}}\text{\hspace{.4mm}}|\text{\hspace{.4mm}}D_{\tilde{e}})$) are then determined by weighting the inferred kick per $\tilde{e}$ as given by Eq.\ \ref{eq8} by the eccentricity distributions and integrating over $\tilde{e}$, or:
\begin{equation}
    \label{eq9}
    P(v_{\text{kick}}\text{\hspace{.4mm}}|\text{\hspace{.4mm}}D_{\tilde{e}})=\int P(v_{\text{kick}}\text{\hspace{.4mm}}|\text{\hspace{.4mm}}\tilde{e})P(\tilde{e}\text{\hspace{.4mm}}|\text{\hspace{.4mm}}D_{\tilde{e}})d\tilde{e}.
\end{equation}
Moreover, when combining the kick estimates of individual pulsars to show the total kick distributions (of pulsars with a certain age), we divide the results by the prior $P(v_{\text{kick}})$ in order to account for the fact that the contribution of highly kicked objects decreases for older pulsars (because of which an old pulsar with a high kick should have an increased weight). We note that this correction (as well as the correction from Fig.\ \ref{fig5}) has no noticeable effect on the results of \citet{Disberg_2024b}.\\
\indent The method described in this section makes several assumptions in its kick estimation. For example, we assumed that pulsar formation takes place in the Galactic disk at $z=0$ kpc, where the initial velocity of the pulsar consists of a perfectly circular Galactic orbit and an isotropic natal kick. Moreover, the relationships shown in Fig.\ \ref{fig4} are produced by integrating our simulation over several time intervals, but the distribution of pulsar ages within these intervals is not uniform. We also do not consider a possible binary origin in which the natal kicks unbind the binary and thus lose a fraction of their initial kinetic energy \citep[e.g.][]{Bailes_1989,Kalogera_1996,Tauris_1998,Kuranov_2009}. However, \citet{Chrimes_2023} find that the velocities of unbound NSs are dominated by the natal kicks, and \citet{Kapil_2023} also conclude that a binary origin has no significant effect on their simulated kicks. \citet{Renzo_2019}, in turn, find that only ${\sim}1\%$ of disrupted massive stars (i.e.\ potential NS progenitors) obtain velocities above $30$ km/s, whereas ejected helium stars can obtain higher velocities but are significantly rarer. The hypothetical effects of a \textquotedblleft rocket\textquotedblright\ mechanism are neglected as well \citep[cf.][]{Hirai_2024}. Moreover, \citet{Spitzer_1951,Spitzer_1953} discuss the effect of overdensities in the Galactic disc (such as giant molecular clouds) on stellar velocities \citep[see also e.g.][]{Dehnen_1998,Sellwood_2002}, and \citet{Madau_1994} argue that this dynamical heating should apply to (old) NSs as well. However, \citet{Ofek_2009} argues that dynamical heating becomes important for velocities below $(\tau_{\text{c}}\cdot328\,\text{km}^{2}\,\text{s}^{-2}\,\text{Gyr}^{-1})^{1/2}$, meaning that for an old pulsar with $\tau_{\text{c}}=1$ Gyr it is relevant when the velocity is less than ${\sim}20$ km/s, and he therefore concludes this effect can be neglected due to the relatively high velocities of NSs. Another potential effect assumed to be negligible is the alignment of kick and NS spin \citep[e.g.][]{Johnston_2005,Biryukov_2024} because of which the observed pulsar velocities are not isotropic and the kicks of young pulsars might be overestimated by ${\sim}15\%$ \citep{Mandel_2023}. These assumptions add uncertainty to the kick estimates that follow from this method, although we note that most of them apply to the method of \citet{Verbunt_2017} and \citet{Igoshev_2020} as well.
\section{Kicks}
\label{sec4}
Having estimated the eccentricities of the pulsars in the Verbunt et al. and Igoshev samples, we used the method of \citet{Disberg_2024b}---as described above---to infer the natal kicks of these pulsars. We show and discuss the resulting kick distributions for both samples and for different ranges of $\tau_{\text{c}}$ (Sect.\ \ref{sec4.1}) and compare them to kinematically constrained estimates using the pulsar sample of \citet{Hobbs_2005} as well as distributions established in literature (Sect.\ \ref{sec4.2}).
\subsection{Results}
\label{sec4.1}
\begin{figure*}
    \centering
    \includegraphics[width=18cm]{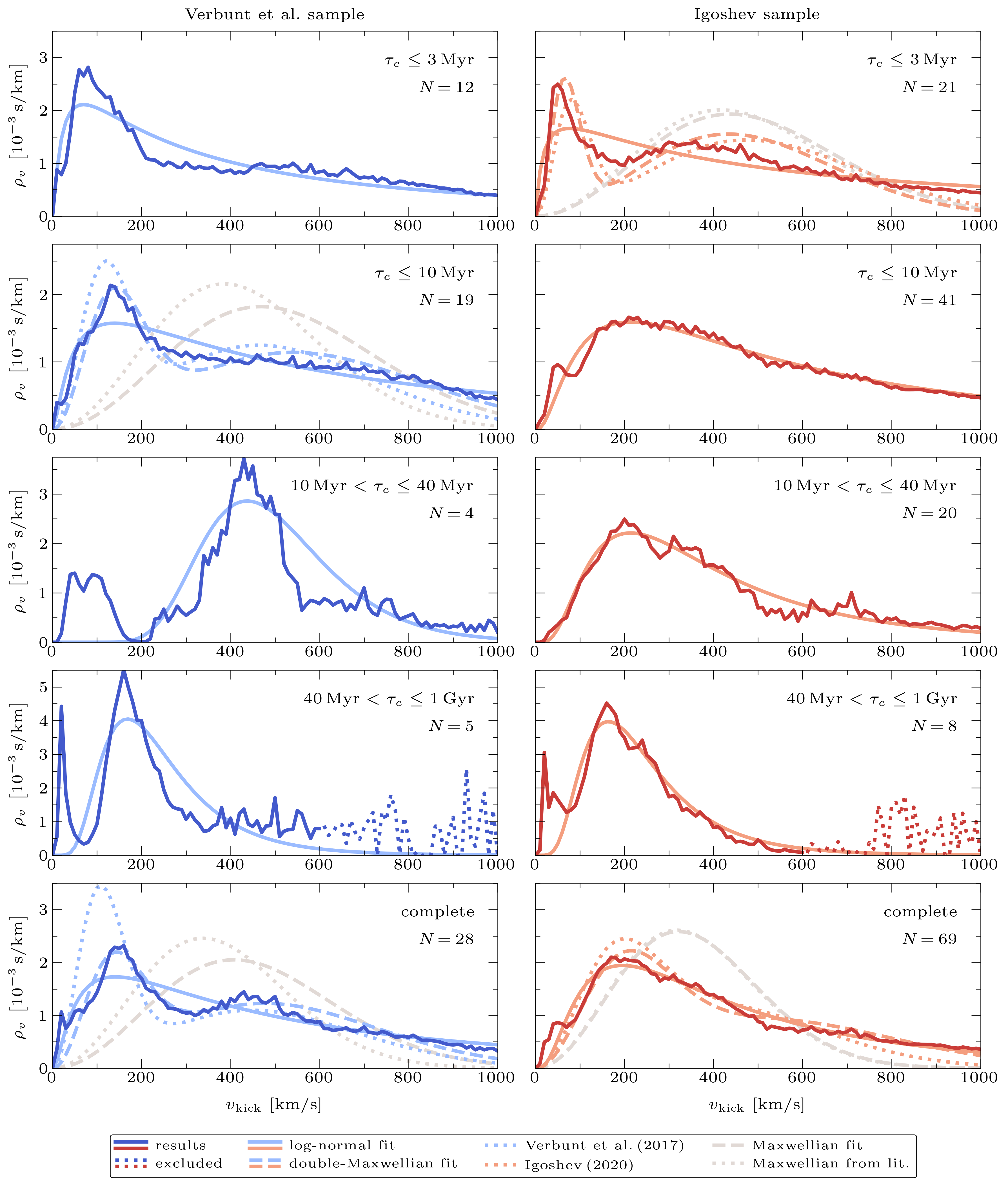}
    \caption{Resulting kick estimates (blue and red solid lines) for the pulsars in the sample of \citet[][blue and left column]{Verbunt_2017} and the expanded sample of \citet[][red and right column]{Igoshev_2020} after applying the method described in Sect.\ \ref{sec3} to the eccentricity estimates shown in Fig.\ \ref{fig3}. The samples---consisting of $N$ pulsars---are divided in four age ranges (i.e.\ $\tau_{\text{c}}\leq3$ Myr, $\tau_{\text{c}}\leq10$ Myr, $10$ Myr $<\tau_{\text{c}}\leq40$ Myr, $40$ Myr $<\tau_{\text{c}}\leq1$ Gyr, and the complete sample) as shown in the different rows. We fitted log-normal distributions (solid light blue/red) to the results in each panel (listed in Table \ref{tab1}), and to the results corresponding to the ages analysed in the literature we also fitted Maxwellians and double-Maxwellians (light blue/red and grey dashed lines, listed in Table \ref{tab2}). The corresponding distributions posed by \citet{Verbunt_2017} and \citet{Igoshev_2020} are also shown (light blue/red and grey dotted lines). For ages $40$ Myr $<\tau_{\text{c}}\leq1$ Gyr we exluded the results for $v_{\text{kick}}>600$ km/s (dotted blue/red lines) in the normalisation and fits. We adopted the log-normal fit to the results for the $\tau_{\text{c}}\leq10$ Myr Igoshev sample as the fiducial natal NS kick distribution.}
    \label{fig6}
\end{figure*}
\begin{table}
\centering
\caption{Parameters of the log-normal distributions (Eq.\ \ref{eq10}) fitted to the results shown in Fig.\ \ref{fig6}, for the pulsar samples of \citet{Verbunt_2017} and \citet{Igoshev_2020}.}
\label{tab1}
\begin{tabular}{lllll}
\hline\hline\\[-10pt]
ages & \multicolumn{2}{c}{\hspace{-1.5mm}Verbunt et al.\ sample} & \multicolumn{2}{c}{\hspace{-1.5mm}Igoshev sample}\\
$\left[\text{Myr}\right]$ & \hspace{4mm}$\mu$ & \hspace{4mm}$\sigma$ & \hspace{4mm}$\mu$ & \hspace{4mm}$\sigma$\\\hline\\[-10pt]
$\tau_c\leq3$ & 6.36(12) & 1.46(6) & 7.38(28) & 1.74(11)\\
$\tau_c\leq10$ & 6.73(10) & 1.34(5) & \textbf{6.38(3)}\tablefootmark{a} & \textbf{1.01(2)}\tablefootmark{a}\\
$10\leq\tau_c\leq40$ & 6.17(2) & 0.31(2) & 5.86(2) & 0.71(1)\\
$40\leq\tau_c\leq10^3$ & 5.39(4) & 0.51(4) & 5.38(3) & 0.53(2)\\
complete & 6.37(8) & 1.19(5) & 6.00(2) & 0.85(2)\\\hline
\end{tabular}
\tablefoot{
Errors equal the variances on the least-squares fits. \tablefoottext{a}{Fiducial kick distribution.}
}
\end{table}
\begin{table*}
\centering
\caption{Parameters of the (double-)Maxwellians fitted to our results for the \citet{Verbunt_2017} and \citet{Igoshev_2020} samples (fit), compared to the results from the corresponding literature (lit.), for young pulsars and the complete samples.}
\label{tab2}
\begin{tabular}{lllllll}
\hline\hline\\[-10pt]
\multicolumn{3}{l}{sample} & \multicolumn{3}{c}{double-Maxwellian\tablefootmark{a}} & Maxwellian\tablefootmark{b}\\
& & & \multicolumn{1}{c}{$\sigma_1\ \left[\text{km/s}\right]$} & \multicolumn{1}{c}{$\sigma_2\ \left[\text{km/s}\right]$} & \multicolumn{1}{c}{$w$} & \multicolumn{1}{c}{$\sigma\ \left[\text{km/s}\right]$}\\\hline\\[-10pt]
\multirow{4}{*}{Verbunt et al.} & \multirow{2}{*}{young ($\tau_c\leq10\,\text{Myr}$)} & fit & \hspace{2mm}96(1) & \hspace{2mm}382(6) & \hspace{2mm}0.31(1) & \hspace{3.5mm}331(13)\\
 & & lit. & \hspace{2mm}82(27) & \hspace{2mm}328(53) & \hspace{2mm}0.32(16) & \hspace{3.5mm}273(33) \\\cline{2-7}\\[-10pt]
 & \multirow{2}{*}{complete} & fit. & \hspace{2mm}97(2) & \hspace{2mm}362(7) & \hspace{2mm}0.31(1) & \hspace{3.5mm}311(11)\\
 & & lit. & \hspace{2mm}75(17) & \hspace{2mm}316(47) & \hspace{2mm}0.42(13) & \hspace{3.5mm}239(24) \\\hline\\[-10pt]
\multirow{4}{*}{Igoshev} & \multirow{2}{*}{young ($\tau_c\leq3\,\text{Myr}$)} & fit & \hspace{2mm}46(1) & \hspace{2mm}307(6) & \hspace{2mm}0.20(1) & \hspace{3.5mm}309(11)\\
 & & lit. & \hspace{2mm}56(10) & \hspace{2mm}336(45) & \hspace{2mm}0.20(11) & \hspace{3.5mm}295(34) \\\cline{2-7}\\[-10pt]
 & \multirow{2}{*}{complete} & fit. & \hspace{2mm}144(3) & \hspace{2mm}400(11) & \hspace{2mm}0.46(1) & \hspace{3.5mm}233(6)\\
 & & lit. & \hspace{2mm}128(20) & \hspace{2mm}298(28) & \hspace{2mm}0.42(16) & \hspace{3.5mm}225(18)\\\hline
\end{tabular}
\tablefoot{
The errors on our fits equal the variances of the least-squares fits to our results, whereas the errors in the literature are $68\%$ confidence intervals. \tablefoottext{a}{Defined in Eq.\ \ref{eq12}.}
\tablefoottext{b}{Defined in Eq.\ \ref{eq11}.}
}
\end{table*}
We applied the method discussed in Sect.\ \ref{sec3} to the eccentricity estimates shown in Sect.\ \ref{sec2}, for the Verbunt et al.\ sample and the expanded Igoshev sample. We determined the results for these complete samples as well as the subsets shown in Fig.\ \ref{fig3} (i.e.\ $\tau_{\text{c}}\leq3$ Myr, $\tau_{\text{c}}\leq10$ Myr, $10$ Myr $<\tau_{\text{c}}\leq40$ Myr, and $40$ Myr $<\tau_{\text{c}}\leq1$ Gyr). In Fig.\ \ref{fig6} we show our resulting kick distributions for the pulsar samples, and fit multiple distributions to the results. That is, similarly to \citet{Disberg_2024b} we fit log-normal distributions:
\begin{equation}
    \label{eq10}
    \rho_{\text{logn}}(v\text{\hspace{.4mm}}|\text{\hspace{.4mm}}\mu,\sigma)=\dfrac{1}{v\,\sigma\sqrt{2\pi}}\exp\left(-\dfrac{\left(\ln v-\mu\right)^2}{2\sigma^2}\right),
\end{equation}
where the fitted parameters are listed in Table \ref{tab1}. Moreover, similarly to \citet{Verbunt_2017} and \citet{Igoshev_2020} we fit Maxwellians to our results (i.e.\ the results for the complete samples and their \textquotedblleft young\textquotedblright\ subsets): 
\begin{equation}
    \label{eq11}
    \rho_{\text{M}}(v\text{\hspace{.4mm}}|\text{\hspace{.4mm}}\sigma)=\sqrt{\dfrac{2}{\pi}}\dfrac{v^2}{\sigma^3}\exp\left(\dfrac{-v^2}{2\sigma^2}\right).
\end{equation}
They find, however, that NS kicks are better described by a distribution that combines two Maxwellians. We therefore also fit a double-Maxwellian to our results, defined as:
\begin{equation}
    \label{eq12}
    \rho_{\text{2M}}(v\text{\hspace{.4mm}}|\text{\hspace{.4mm}}\sigma_1,\sigma_2,w)=w\cdot\rho_{\text{M}}(v\text{\hspace{.4mm}}|\text{\hspace{.4mm}}\sigma_1)+(1-w)\cdot\rho_{\text{M}}(v\text{\hspace{.4mm}}|\text{\hspace{.4mm}}\sigma_2),
\end{equation}
where $\sigma_1\leq\sigma_2$ and the parameter $w$ determines the fraction of NS kicks situated in the lower Maxwellian component. The distributions of our results as well as the fitted distributions are normalised between $0$ km/s and $1000$ km/s. The fitted parameters of the Maxwellians and double-Maxwellians are listed in Table \ref{tab2} and compared to the parameters from the corresponding literature.\\
\indent The first two rows in Fig.\ \ref{fig6} respectively show the estimated kick distributions for pulsars younger than $3$ Myr \citep[similar to][]{Hobbs_2005,Igoshev_2020} and younger than $10$ Myr \citep[similar to][]{Verbunt_2017}. When we fit (double-)Maxwellians to the results corresponding to the \textquotedblleft young\textquotedblright\ subsets of the Verbunt et al.\ and the Igoshev samples, we find that the resulting distributions resemble the corresponding fits of \citet{Verbunt_2017} and \citet{Igoshev_2020} remarkably well. We thus find that our kick distributions for the young pulsars in their samples agree with their results. The single Maxwellian fits, however, are clearly not representative of the underlying kick distribution, because of which a comparison with, for instance, the Maxwellian found by \citet{Hobbs_2005} is not meaningful. The double-Maxwellian fits, in contrast, are an accurate description of the results. However, although the kicks of the $\tau_{\text{c}}\leq3$ Myr pulsars in the Igoshev sample show a clear bimodality, the kicks of the corresponding $\tau_{\text{c}}\leq10$ Myr pulsars in the Verbunt et al.\ sample are less bimodal than the double-Maxwellian fit suggests. This might be due to our method not being sensitive enough to resolve the bimodal structure, but since \citet{Verbunt_2017} determine their fits through a direct comparison with observed data there is no way to examine whether they actually do find a clearly bimodal structure. Nevertheless, our results do suggest a bimodality for the $\tau_{\text{c}}\leq3$ Myr pulsars in the Igoshev sample, and are compatible with bimodality in the $\tau_{\text{c}}\leq10$ Myr pulsars in the Verbunt et al. sample.\\
\indent However, the $\tau_{\text{c}}\leq10$ Myr pulsars in the Igoshev sample do not show a bimodal structure at all (in fact, the kicks agree well with the log-normal fit). This set of pulsars contains the \textquotedblleft young\textquotedblright\ subsets of both the Verbunt et al.\ and the Igoshev samples, so if the bimodality of those double-Maxwellian fits were physical then one would expect the $\tau_{\text{c}}\leq10$ Myr pulsars in the Igoshev sample to also follow a bimodal distribution. These results, though, show a single peak at $v_{\text{kick}}\approx200$ km/s at the same place where the $\tau_{\text{c}}\leq3$ Myr subset shows a dearth of pulsars. This is difficult to explain if the true natal kick distribution for NSs follows a bimodal distribution similar to the one found by \citet{Igoshev_2020}. Indeed, in the subsequent age ranges the older pulsars in the Igoshev sample also follow an approximately log-normal distribution peaking at $v_{\text{kick}}\approx200$ km/s. The corresponding distributions for the Verbunt et al.\ sample only contain $4$ and $5$ pulsars, respectively, because of which it is difficult to draw strong conclusions from these results. For the oldest pulsars the correction of dividing by the prior kick probabilities causes unnecessary noise at $v_{\text{kick}}>600$ km/s, which is why we exclude this region from the fits and normalisations.\\
\begin{figure}
    \resizebox{\hsize}{!}{\includegraphics{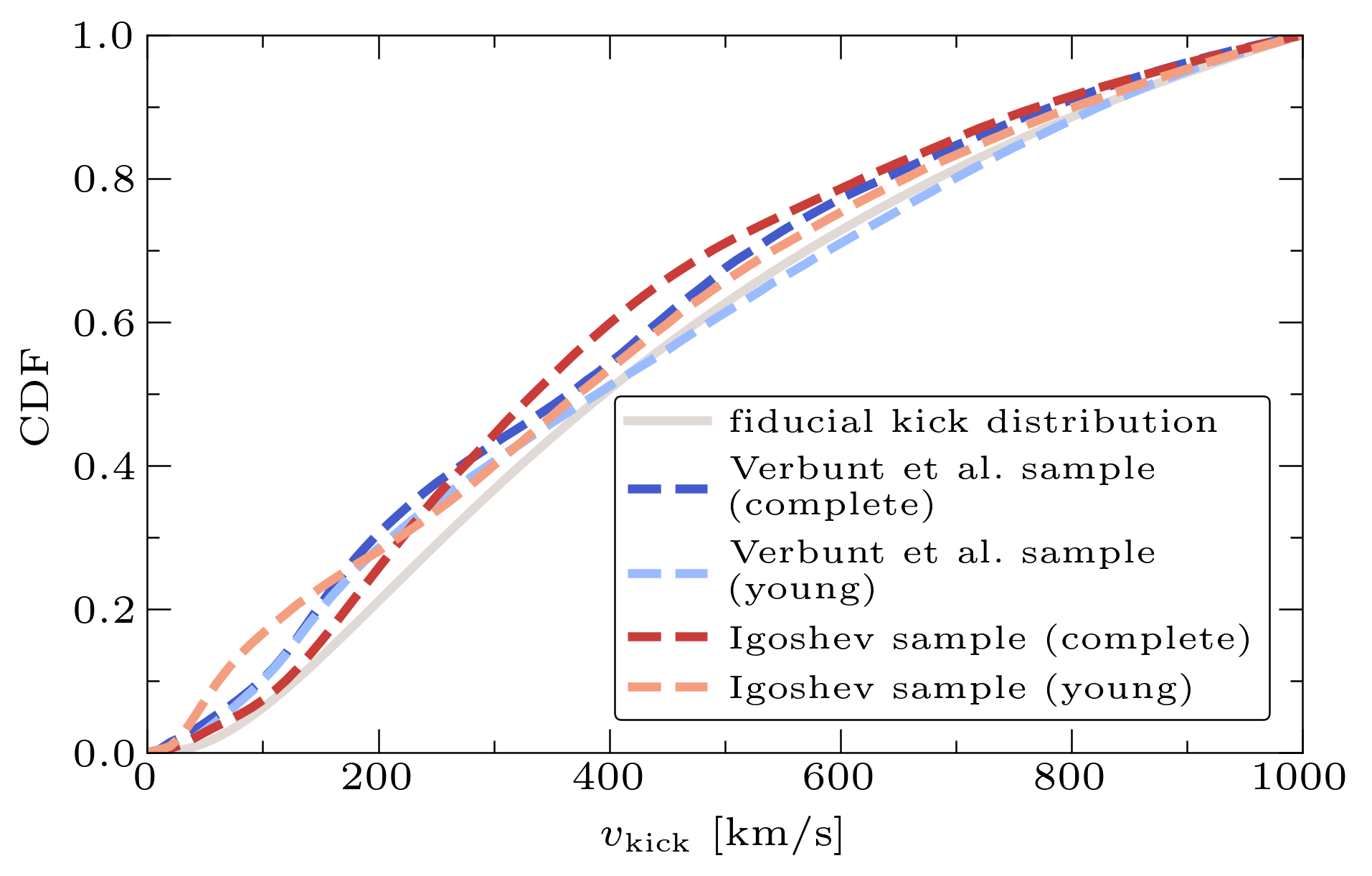}}
    \caption{Cumulative distributions of the fiducial kick distribution (log-normal with $\mu=6.00$ and $\sigma=0.85$, grey) compared to the kick distributions from \citet[][blue]{Verbunt_2017} and \citet[][red]{Igoshev_2020}, for their complete samples as well as young subsets (blue/red and light blue/red, respectively). The distributions are all normalised in the region of $0$ km/s $\leq v_{\text{kick}}\leq1000$ km/s.}
    \label{fig7}
\end{figure}
\indent In the bottom row of Fig.\ \ref{fig6} we show the kick distributions for the complete samples (i.e.\ the sum of the separate distributions weighted by the amount of pulsars). Our results for the complete Verbunt et al.\ sample differ slightly from the fit made by \citet{Verbunt_2017}. This is likely due to the fact that we find that this sample contains relatively high kicks for pulsars older than $10$ Myr, and \citet{Disberg_2024a} show that---up to a certain point---higher kicks result in lower observed velocities for older pulsars. Therefore the older pulsars probably contribute to the lower Maxwellian peak of \citet{Verbunt_2017}, while we find that this is actually an underestimation of their actual kick. Our results for the complete Verbunt et al.\ sample do, however, match the results for the complete Igoshev sample better. In fact, the fit made by \citet{Igoshev_2020} also aligns with our results (and is less bimodal than the double-Maxwellian structure might suggest). This agreement is probably coincidental, since the (kick-independent) velocity distribution of older pulsars found by \citet{Disberg_2024a} might be aligning with the natal kick distribution (peaking at ${\sim}200$ km/s). Nevertheless, we argue that (1) the kinematically constrained pulsar kick distributions are consistent with a single peak at $200$ km/s, and (2) the resulting kick distribution for the complete Igoshev sample appears to describe the distributions for the different age-selected samples relatively well and approximately follows a log-normal distribution.\\
\indent The question arises whether the bimodality in the young pulsar kicks is consistent with the log-normal distributions that older pulsars appear to follow. Considering that (relatively) young pulsars remain the least biased with regard to kick velocities, we adopted the log-normal fit to the Igoshev sample for $\tau_{\text{c}}\leq10$ Myr (i.e.\ $\mu=6.38$ and $\sigma=1.01$) as the fiducial NS natal kick distribution determined through kinematic constraints. Furthermore, we employed a one-sample Kolmogorov-Smirnov (KS) test in order to evaluate the hypothesis that the underlying natal kick distribution for the results in the different age ranges equals the fiducial log-normal distribution. Relevantly, in Fig.\ \ref{fig7} we show the cumulative distribution functions (CDFs) of the young and complete samples as examples. We computed the KS statistic for the results shown in the panels of Fig.\ \ref{fig6}, and find that for no distribution it exceeds the critical value \citep[as listed by][]{Massey_1951} above which we can reject the hypothesis at a $0.05$ level of significance (although for $\tau_{\text{c}}>40$ Myr the difference becomes relatively small). We therefore conjecture that---despite the fact that \citet{Valli_2025} predict that unbound NSs indeed form a second peak at ${\sim}100$ km/s---this bimodality may be caused by noise due to low-number statistics (cf.\ the analysis of the observed binary black hole mass distribution of \citeauthor{Farah_2023} \citeyear{Farah_2023} and \citeauthor{Adamcewicz_2024} \citeyear{Adamcewicz_2024}). Moreover, one might suspect that there are pulsars in the young samples that are older than their $\tau_{\text{c}}$ might suggest. We therefore investigated the results for the young samples where we excluded the pulsars with a $\tau_{\text{kin}}$ that exceeds $\tau_{\text{c}}$ by more than $5$ Myr (as marked in Fig.\ \ref{fig2}). In Appendix \ref{App.D} we show that these results still include the bimodality, meaning it is probably not caused by pulsars of which $\tau_{\text{kin}}$ and $\tau_{\text{c}}$ do not align.
\subsection{Comparisons}
\label{sec4.2}
\begin{figure}
    \resizebox{\hsize}{!}{\includegraphics{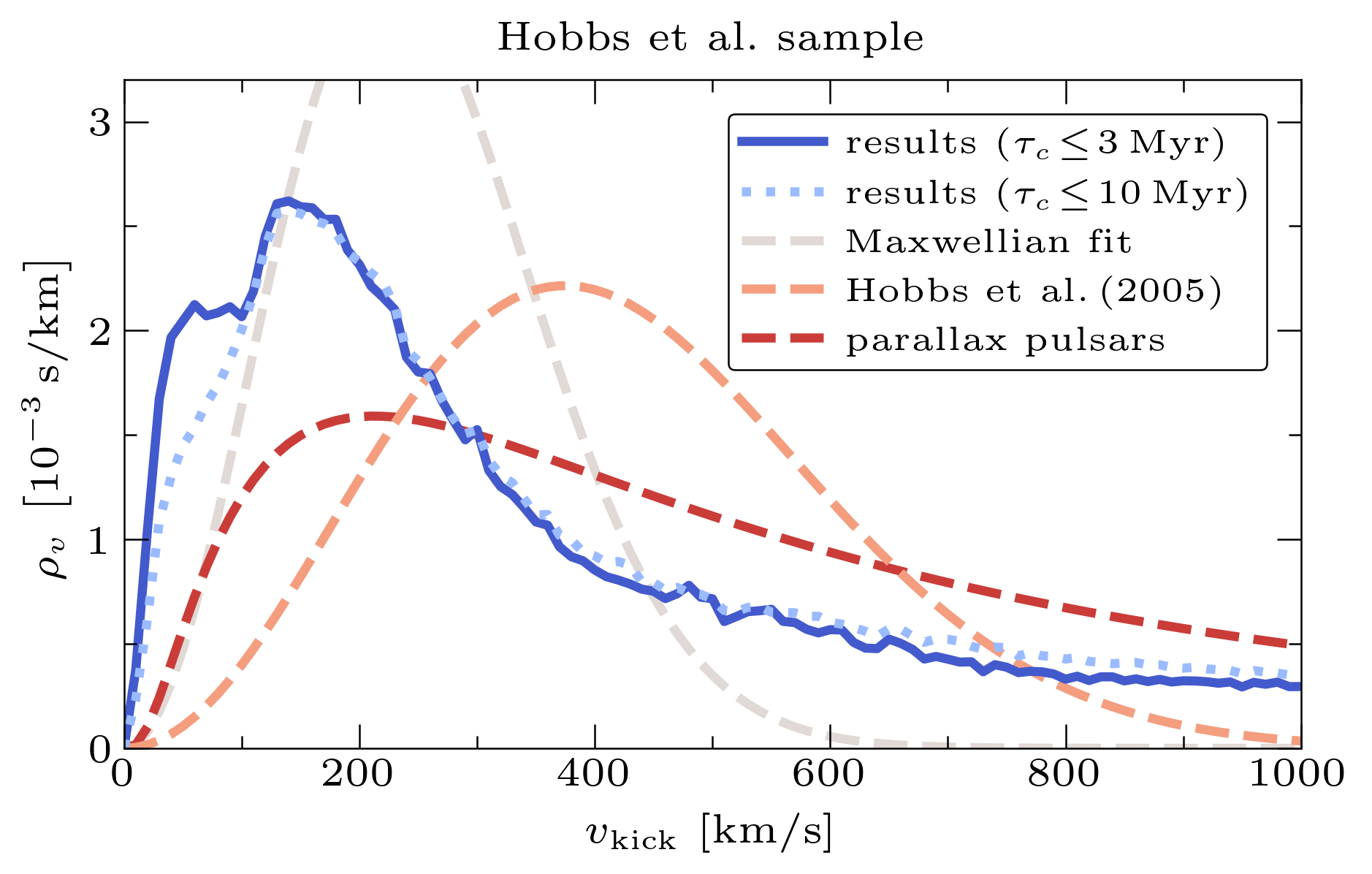}}
    \caption{Kinematically constrained kick distributions for the young pulsars using the pulsar sample---including distance estimates---of \citet{Hobbs_2005}, considering the age ranges of $\tau_{\text{c}}\leq3$ Myr (blue solid line, with a sample size $N=46$) and $\tau_{\text{c}}\leq10$ Myr (light blue dotted line, $N=83$). The grey dashed line shows a Maxwellian fit to the $\tau_{\text{c}}\leq3$ Myr results (resulting in $\sigma=159(5)$ km/s). For comparison we also show the fit of \citet[][orange dashed line, a Maxwellian with $\sigma=265$ km/s]{Hobbs_2005} and the fiducial kick distribution based on the results for the pulsars with parallax estimates (red dashed line).}
    \label{fig8}
\end{figure}
\begin{figure*}
    \centering
    \includegraphics[width=18cm]{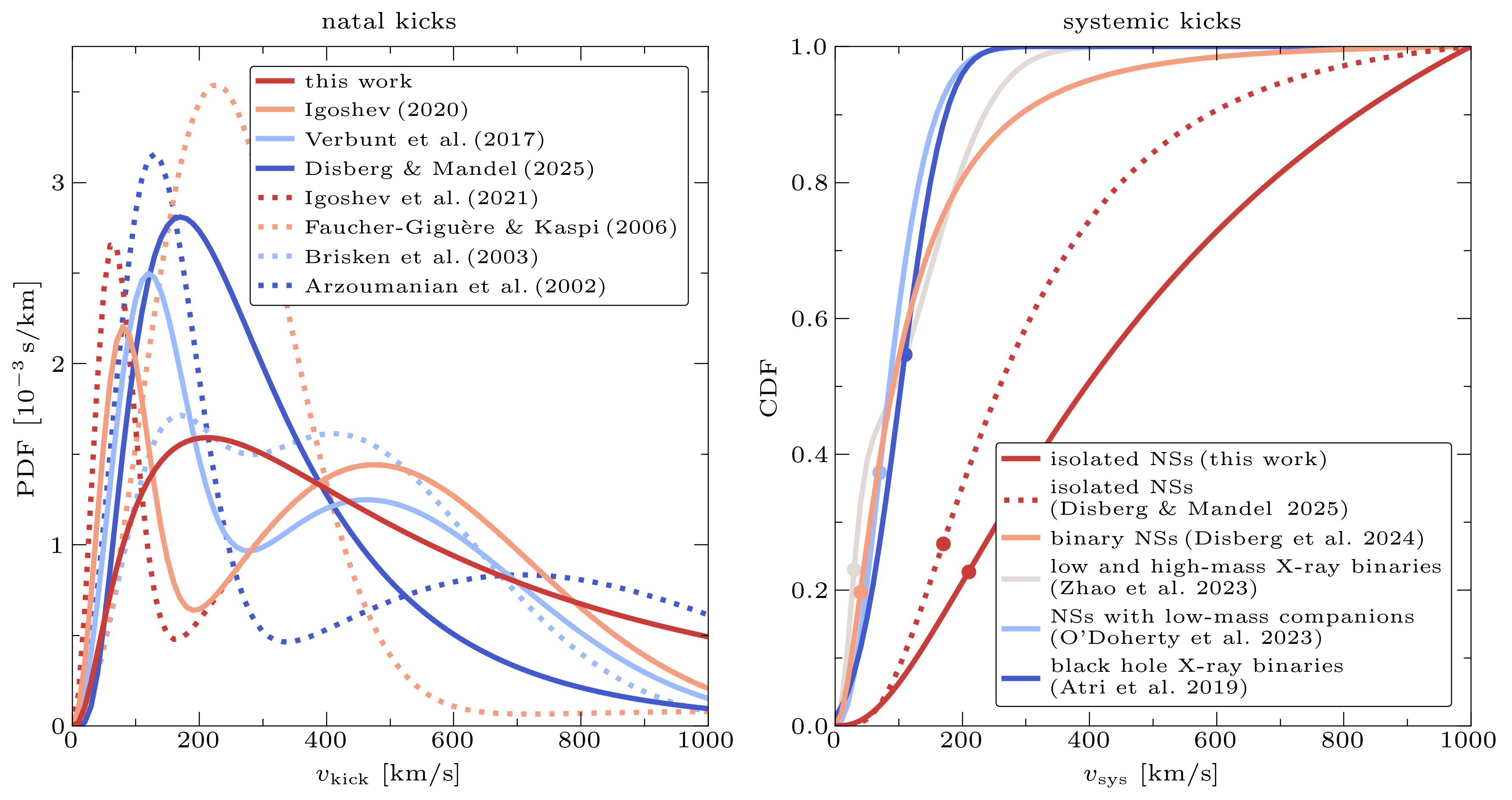}
    \caption{Comparison between our fiducial kick distribution (red solid line) and kick distributions as found in existing literature. In the left panel the comparison is made with probability density functions (PDFs) from existing literature on NS natal kicks \citep[i.e.][omitting the erroneous distribution of \citeauthor{Hobbs_2005} \citeyear{Hobbs_2005}]{Arzoumanian_2002,Brisken_2003,Faucher_2006,Verbunt_2017,Igoshev_2020,Igoshev_2021,Disberg_2025}, while in the right panel we compare our natal kicks and the results of \citet{Disberg_2025} to CDFs of the systemic kicks of binaries that include compact objects \citep[i.e.][where the modes are displayed as dots]{Atri_2019,ODoherty_2023,Zhao_2023,Disberg_2024b}. The shown distributions are all normalised between $0$ km/s and $1000$ km/s, and their parameters are listed in Appendix \ref{App.A}.}
    \label{fig9}
\end{figure*}
The fiducial kick distribution and the distributions found by \citet{Verbunt_2017} and \citet{Igoshev_2020} contain lower kick velocities than the commonly used kick distribution of \citet[][i.e.\ a Maxwellian with $\sigma=265$ km/s]{Hobbs_2005}. This difference could hypothetically be caused by the fact that the Hobbs et al.\ sample is not limited to pulsars with observed parallaxes but mainly uses distance estimates through dispersion measures. In order to test this, we applied our method to the Hobbs et al.\ sample---using their distance estimates---and compared it to the parallax samples from Fig.\ \ref{fig6}. In Fig.\ \ref{fig8} we show the results for the pulsars in the Hobbs et al.\ sample that are either younger than $3$ Myr \citep[for comparison with][]{Hobbs_2005,Igoshev_2020}
or $10$ Myr \citep[for comparison with][]{Verbunt_2017}. Both of these age ranges result in a kick distribution showing one clear peak at ${\sim}100{-}200$ km/s.\\
\indent The figure shows how our results differ significantly from the Maxwellian found by \citet{Hobbs_2005}, which peaks at ${\sim}400$ km/s. In Fig.\ \ref{fig6} the single-Maxwellian fits peak at these high kick velocities as well, but (1) this is coincidental since \citet{Hobbs_2005} show that their results follow their fit well (in contrast to the single Maxwellians in Fig.\ \ref{fig6}), and (2) the Maxwellian fit in Fig.\ \ref{fig8} differs significantly. More importantly, our results for the Hobbs et al.\ sample are more similar to our results for the parallax samples, and seem relatively well-described by our fiducial kick distribution. The latter includes slightly higher kicks, though, but this is similar to the difference between the Verbunt et al.\ sample and the Igoshev sample. In fact, we note that the results for the young pulsars in the Hobbs et al.\ sample (Fig.\ \ref{fig8}) are especially similar to the results for the young pulsars in the Verbunt et al.\ sample (Fig.\ \ref{fig6}). We therefore argue that it may be difficult to explain the difference between the results of these studies through inaccuracy in distance estimates. After all, we were able to use our method to recreate the results of \citet{Verbunt_2017} and \citet{Igoshev_2020}, but failed to reproduce the results of \citet{Hobbs_2005} even though we used their distance estimates. Lastly, we note the similarity between our kick estimates (in particular the results for the young pulsars in the Hobbs et al.\ and Verbunt et al.\ samples) and the theoretical---probabilistic---kick prescription of \citet[][see also \citeauthor{Mandel_2021} \citeyear{Mandel_2021} and \citeauthor{Kapil_2023} \citeyear{Kapil_2023}]{Mandel_2020}.\\
\indent While this article was in review, \citet{Disberg_2025} analysed the present-day velocities of young pulsars in order to obtain a kick velocity distribution (and applied the method described in this paper to pulsars older than $10$ Myr, finding consistent results). Their resulting kick distribution is in general similar to our fiducial kick distribution: it is a log-normal distribution peaking at ${\sim}200$ km/s. The main difference between these two distributions is that they predict a smaller contribution of the high-velocity tail. After all, as stated before, our method is mainly sensitive to velocities $\leq400$ km/s. Moreover, \citet{Disberg_2025} show that the Maxwellian distribution found by \citet{Hobbs_2005} is based on an erroneous histogram interpretation that misses a Jacobian to correct for its logarithmic bin sizes. Correcting for the missing Jacobian indeed results in kicks similar to our results shown in Fig.\ \ref{fig8}, explaining why we are unable to reproduce the Maxwellian distribution of \citet{Hobbs_2005} and solving the tension in the literature caused by their work.\\
\indent In the left panel of Fig.\ \ref{fig9} we compare our fiducial kick distribution to results from literature \citep[i.e.][]{Arzoumanian_2002,Brisken_2003,Faucher_2006,Verbunt_2017,Igoshev_2020,Igoshev_2021,Disberg_2025}. The figure shows that existing literature on NS kicks indeed tend to find distributions that peak around ${\sim}100{-}200$ km/s. Since we omit the distribution of \citet{Hobbs_2005}, the literature paints a relatively consistent picture. In the right panel of Fig.\ \ref{fig9} we also show the results of studies estimating the systemic kicks of binaries that include a NS or black hole \citep[i.e.][]{Atri_2019,ODoherty_2023,Zhao_2023,Disberg_2024b}. The natal kicks of isolated NSs peak at kick velocities a factor ${\sim}2$ higher than the systemic kicks of these binaries. There are at least two reasons for this: (1) high natal kicks can unbind a binary, hence the observed binaries that experienced a SN kick are biased towards low kicks \citep[e.g.][]{Beniamini_2016}, and (2) the natal kick contributes to the systemic kick with a weight equal to the NS mass over the binary mass, and therefore the systemic kicks is usually smaller when the natal kick dominates over the Blaauw kick. In order to investigate these two effects it would be interesting to attempt to reconcile the observed natal and systemic kick distributions, for example through a population synthesis.
\section{Conclusions}
\label{sec5}
In this work, we have analysed the pulsars in the samples of \citet{Verbunt_2017} and \citet{Igoshev_2020}, and estimated the eccentricities of their Galactic orbits (Sect.\ \ref{sec2}). Moreover, we expanded the simulation of \citet{Disberg_2024b} which determines the relationship between kick magnitude and Galactic eccentricity (Sect.\ \ref{sec3}). The results of this simulation were combined with the eccentricity estimates in order to determine the natal kicks of the pulsars in the Verbunt et al.\ and Igoshev samples (Sect.\ \ref{sec4}). Based on our analysis we come to the following conclusions:
\begin{itemize}
\item For most young pulsars in the Verbunt et al.\ and Igoshev samples (Fig.\ \ref{fig1}), their kinematic age and characteristic age align relatively well (Fig.\ \ref{fig2}). This means that LSR isotropy is compatible with the characteristic age estimates, which provides confidence in both assumptions.
\item The younger pulsars follow more eccentric Galactic orbits (Eq.\ \ref{eq6} and Fig.\ \ref{fig3}), which is likely due to the selection effect of older pulsars with high kicks reaching relatively large offsets, making them less likely to be observed. The simulated orbits (Fig.\ \ref{fig4}) follow this trend as well, and display a close relationship between kick magnitude and Galactic eccentricity (at least for $v_{\text{kick}}\lesssim400$ km/s). After correcting the normalisation for the most eccentric orbits (Fig.\ \ref{fig5}), we used this relationship to infer the kick velocities of the pulsars.
\item The resulting kinematically constrained kick distributions, which were divided in different age bins (Fig.\ \ref{fig6}), are consistent with NS natal kicks following a log-normal distribution with $\mu=6.38$ and $\sigma=1.01$---normalised between $0$ and $1000$ km/s---which peaks at ${\sim}200$ km/s and has a median of ${\sim}400$ km/s (Fig.\ \ref{fig7}). In particular, double-Maxwellians (Eq.\ \ref{eq12}) fitted to the results for young pulsars resemble the results of \citet[][for $\tau_{\text{c}}\leq10$ Myr]{Verbunt_2017} and \citet[][for $\tau_{\text{c}}\leq3$ Myr]{Igoshev_2020}. However, we argue that the bimodality found for young pulsar kicks might not be physical but instead caused by noise due to low-number statistics.
\item If we apply our method to the pulsar sample---including distance estimates---of \citet{Hobbs_2005}, the results do not resemble their Maxwellian fit (Fig.\ \ref{fig8}). Instead, our results do not show a significant difference between the kicks of the pulsars in the Hobbs et al.\ sample and ones in the Verbunt et al.\ sample. This can be explained by the fact that \citet{Disberg_2025} show that the Maxwellian distribution of \citet{Hobbs_2005} is based on an erroneous histogram interpretation that misses a Jacobian. Neglecting their Maxwellian, the existing literature on NS kicks paints a relatively consistent picture with distributions peaking at ${\sim}100{-}200$ km/s (Fig.\ \ref{fig9}).
\end{itemize}
In general, we also note that \citet{Disberg_2024b} show that their method gives results very similar to the method of \citet[][who trace back trajectories and analyse peculiar velocities at disc crossings]{Atri_2019}, while we show that this method also gives results similar to the method of \citet{Verbunt_2017}. Because of this we can be relatively confident in these methods.\\
\indent Our fiducial log-normal kick distribution has median and mean values of ${\sim}400$ km/s and ${\sim}600$ km/s, respectively. This is, for example, compatible with the constraint posed by \citet{Popov_2000} that the average NS kick should exceed ${\sim}200{-}300$ km/s. \citet{Willcox_2021}, in turn, estimate that in their sample $5{\pm}2\%$ of isolated NS kick velocities are below $50$ km/s, and they show that in their model this is compatible with observations of binary NSs and NSs in globular clusters \citep[since these have escape velocities on this scale, see e.g.][]{Katz_1975,Sigurdsson_1995}. However, our log-normal fit predicts ${<}1\%$ of kicks to be below this limit, which might be in tension with their model. Nevertheless, in the kinematically constrained kick estimates for the complete Igoshev sample (Fig.\ \ref{fig6}), ${\sim}3\%$ of kicks are below $50$ km/s, which agrees with the findings of \citet{Willcox_2021} and might indicate that our log-normal fit might be insufficient in describing the low end of the kick distribution in detail (e.g.\ in the context of NS populations in globular clusters). These kind of details and further applications of the NS natal kick distribution could provide an interesting basis for future research.

\begin{acknowledgements}
    The authors are thankful to the referee for helpful comments that helped improve this paper, and thank Ilya Mandel for useful discussions. P.D.\ acknowledges support from the Australian Research Council Centre of Excellence for Gravitational Wave Discovery (OzGrav) through project number CE230100016, and thanks Ilya Mandel for useful discussions. N.G.\ acknowledges studentship support from the Dutch Research Council (NWO) under the project number 680.92.18.02, and A.J.L.\ was supported by the European Research Council (ERC) under the European Union’s Horizon 2020 research and innovation programme (grant agreement No. 725246). In this work, we made use of \lstinline{NUMPY} \citep{Harris_2020}, \lstinline{SCIPY} \citep{Virtanen_2020}, \lstinline{MATPLOTLIB} \citep{Hunter_2007}, \lstinline{GALPY} \citep{Bovy_2015}, and \lstinline{ASTROPY}, a community-developed core Python package and an ecosystem of tools and resources for astronomy \citep{Astropy_2013,Astropy_2018,Astropy_2022}.
\end{acknowledgements}
\bibliographystyle{TeXnical/aa_url}
\bibliography{References}

\begin{appendix}
\section{Kick distributions}
\label{App.A}
In Fig.\ \ref{fig9} we display various distributions, of which the parameters are listed in Table \ref{tabA}. For instance, \citet{Arzoumanian_2002} define a two-component Gaussian:
\begin{equation}
    \label{eqA1}
    \rho_{2\text{G}}(v\text{\hspace{.4mm}}|\text{\hspace{.4mm}}\sigma_1,\sigma_2,w)=4\pi v^2\left[w\,\rho_{\text{G}}(v\text{\hspace{.4mm}}|\text{\hspace{.4mm}}\sigma_1)+(1-w)\,\rho_{\text{G}}(v\text{\hspace{.4mm}}|\text{\hspace{.4mm}}\sigma_2)\right],
\end{equation}
using the Gaussian-like function:
\begin{equation}
    \label{eqA2}
    \rho_{\text{G}}(v\text{\hspace{.4mm}}|\text{\hspace{.4mm}}\sigma)=\dfrac{1}{\left(2\pi\sigma^2\right)^{3/2}}\exp\left(\dfrac{-v^2}{2\sigma^2}\right).
\end{equation}
Moreover, \citet{ODoherty_2023} employ a beta distribution:
\begin{equation}
    \label{eqA3}
    \rho_{\beta}(v\text{\hspace{.4mm}}|\text{\hspace{.4mm}}\alpha,\beta,s)=\dfrac{1}{s}\dfrac{\Gamma(\alpha+\beta)}{\Gamma(\alpha)\Gamma(\beta)}\left(\dfrac{v}{s}\right)^{\alpha-1}\left(1-\dfrac{v}{s}\right)^{\beta-1},
\end{equation}
where $\Gamma$ denotes the gamma function and $s$ is a scaling factor above which the function is set to zero.
\vfill
\begin{table}[h]
\centering
\caption{Parameters of kick distributions (see Fig.\ \ref{fig9}).}
\label{tabA}
\begin{tabular}{ll}
\hline\hline\\[-10pt]
reference & parameter\\\hline\\[-10pt]
this work (INSs)\tablefootmark{a}\tablefootmark{$\dagger$} & $\mu=6.38$\\
 & $\sigma=1.01$\\\hline\\[-10pt]
\citet[][INSs]{Disberg_2025}\tablefootmark{a} & $\mu=5.60$\\
 & $\sigma=0.68$\\\hline\\[-10pt]
\citet[][BNSs]{Disberg_2024b}\tablefootmark{a} & $\mu=4.57$\\
 & $\sigma=0.89$\\\hline\\[-10pt]
\citet[][LMBs]{ODoherty_2023}\tablefootmark{b} & $\alpha=3.05$\\
 & $\beta=14.6$\\
 & $s=563\ \text{km/s}$\\\hline\\[-10pt]
\citet[][XRBs]{Zhao_2023}\tablefootmark{c} & $\sigma_1=21.3\ \text{km/s}$\\
 & $\sigma_2=106.7\ \text{km/s}$\\
 & $w=0.4$\\\hline\\[-10pt]
\citet[][INSs]{Igoshev_2021}\tablefootmark{c} & $\sigma_1=45\ \text{km/s}$\\
 & $\sigma_2=336\ \text{km/s}$\\
 & $w=0.20$\\\hline\\[-10pt]
\citet[][INSs]{Igoshev_2020}\tablefootmark{c} & $\sigma_1=56\ \text{km/s}$\\
 & $\sigma_2=336\ \text{km/s}$\\
 & $w=0.20$\\\hline\\[-10pt]
\citet[][XRBs]{Atri_2019}\tablefootmark{d} & $\mu=107\ \text{km/s}$\\
 & $\sigma=56\ \text{km/s}$\\\hline\\[-10pt]
\citet[][INSs]{Verbunt_2017}\tablefootmark{c} & $\sigma_1=82\ \text{km/s}$\\
 & $\sigma_2=328\ \text{km/s}$\\
 & $w=0.32$\\\hline\\[-10pt]
\citet[][INSs]{Faucher_2006}\tablefootmark{e} & $\sigma_1=160\ \text{km/s}$\\
 & $\sigma_2=780\ \text{km/s}$\\
 & $w=0.90$\\\hline\\[-10pt]
\citet[][INSs]{Hobbs_2005}\tablefootmark{f}\tablefootmark{$\ddagger$} & $\sigma=265\ \text{km/s}$\\\hline\\[-10pt]
\citet[][INSs]{Brisken_2003}\tablefootmark{e} & $\sigma_1=99\ \text{km/s}$\\
 & $\sigma_2=294\ \text{km/s}$\\
 & $w=0.20$\\\hline\\[-10pt]
\citet[][INSs]{Arzoumanian_2002}\tablefootmark{e} & $\sigma_1=90\ \text{km/s}$\\
 & $\sigma_2=500\ \text{km/s}$\\
 & $w=0.40$
\end{tabular}
\tablefoot{
These distributions are either valid for isolated NSs (INSs), binary NSs (BNSs), NSs with a low-mass companion (LMBs), or X-ray binaries (XRBs). \tablefoottext{a}{Log-normal distribution (Eq.\ \ref{eq10}).}
\tablefoottext{b}{Beta distribution (Eq.\ \ref{eqA3}).}
\tablefoottext{c}{Double-Maxwellian (Eq.\ \ref{eq12}).}
\tablefoottext{d}{Gaussian.}
\tablefoottext{e}{Two-component Gaussian (Eq.\ \ref{eqA1}).}
\tablefoottext{f}{Maxwellian (Eq.\ \ref{eq11}).} \tablefoottext{$\dagger$}{Normalised between $0$ and $1000$ km/s.} \tablefoottext{$\ddagger$}{Disputed by \citet{Disberg_2025}.}
}
\end{table}
\section{Pulsar distances}
\label{app.B}
\begin{figure}[h]
    \resizebox{\hsize}{!}{\includegraphics{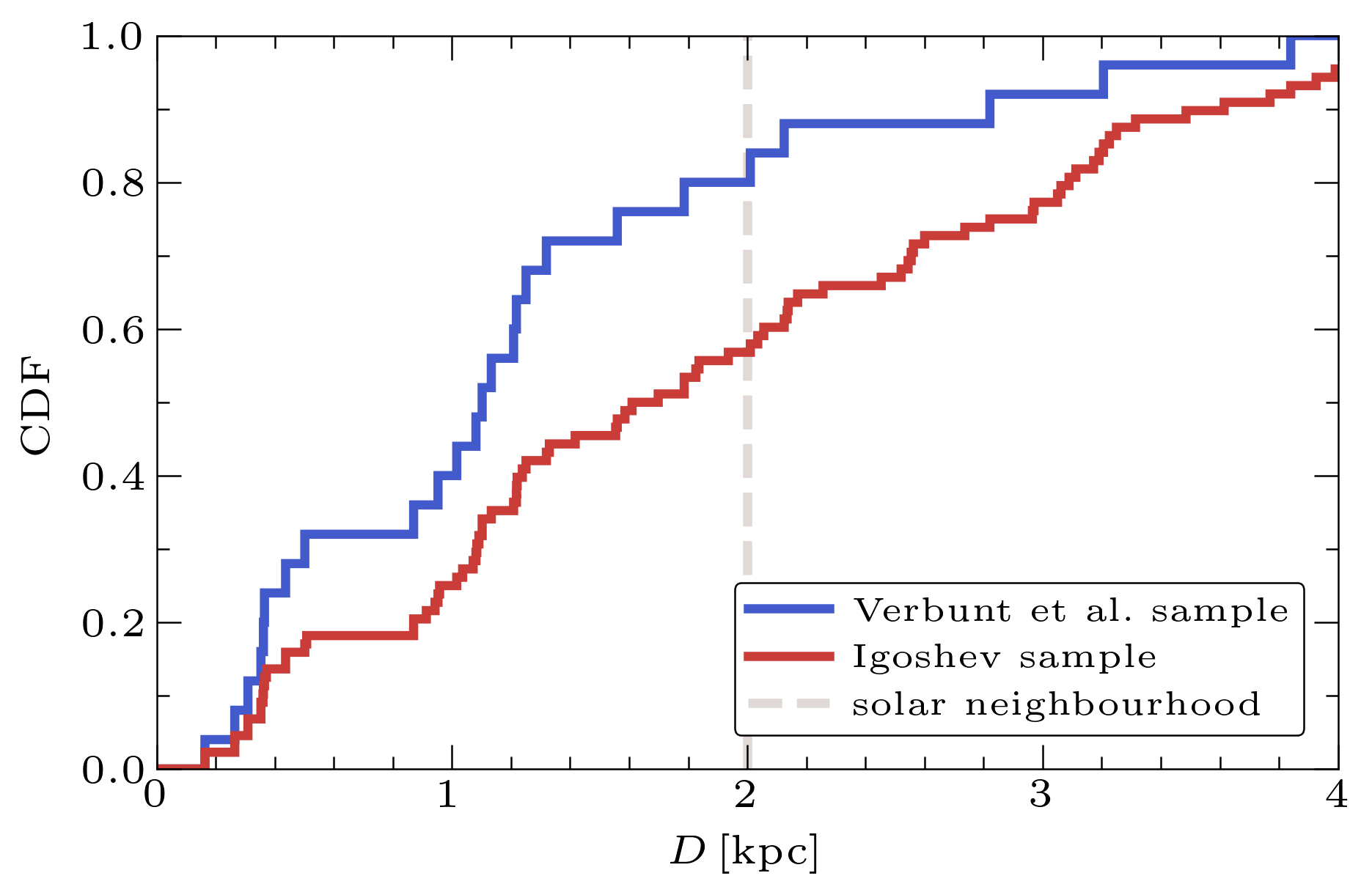}}
    \caption{Cumulative distributions of the pulsar distances corresponding to the samples of \citet[][blue]{Verbunt_2017} and \citet[][red]{Igoshev_2020}, together with the edge of the solar neighbourhood (grey dashed line).}
    \label{figB}
\end{figure}

The distributions of the distances between the Solar System and the pulsars in the samples of \citet{Verbunt_2017} and \citet{Igoshev_2020} are shown in Fig.\ \ref{figB} (for their Galactocentric locations, see Fig.\ \ref{fig1}). We note that for ${\gtrsim}50\%$ of pulsars $D\leq2$ kpc.
\section{Total population}
\label{App.C}
\begin{figure}[h]
    \resizebox{\hsize}{!}{\includegraphics{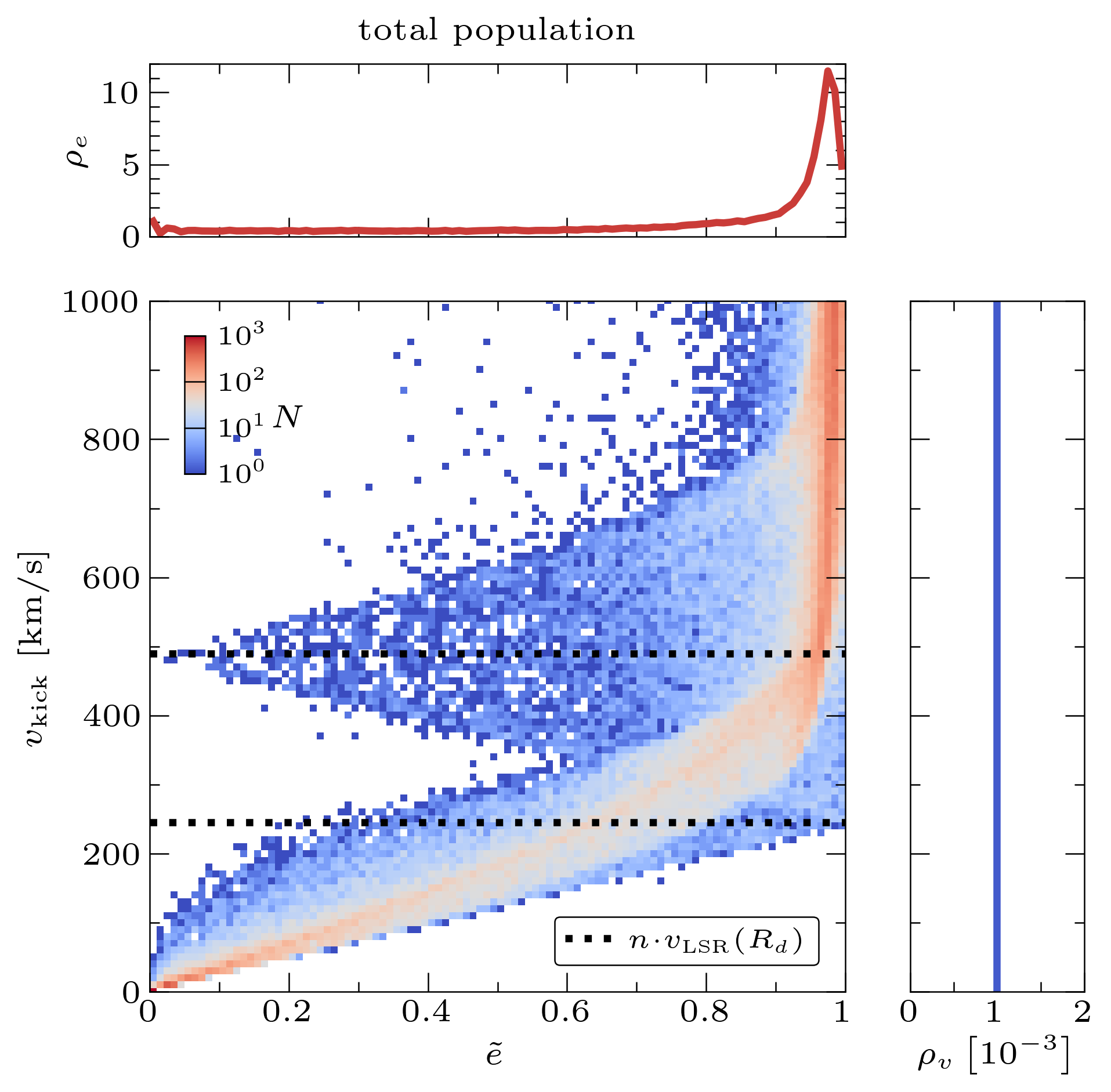}}
    \caption{Distributions of $v_{\text{kick}}$ versus $\tilde{e}$ for the total population as resulting from the simulations (similarly to Fig.\ \ref{fig4}). The distribution also shows the integrated $\tilde{e}$ and $v_{\text{kick}}$ densities ($\rho_e$ and $\rho_v$, respectively). The black dotted lines show one and two times the circular velocity at $R_d=7.04$ kpc (from Eq.\ \ref{eq7}).}
    \label{figC}
\end{figure}

In Fig.\ \ref{fig4} we show the relationship between kick magnitude and eccentricity for objects that are at a certain time in the solar neighbourhood. In Fig.\ \ref{figC}, we also show the (time-independent) distribution for the total population, which aligns relatively well with the results for the solar neighbourhood.

\renewcommand{\thefigure}{D.1}
\begin{figure*}
    \centering
    \includegraphics[width=18cm]{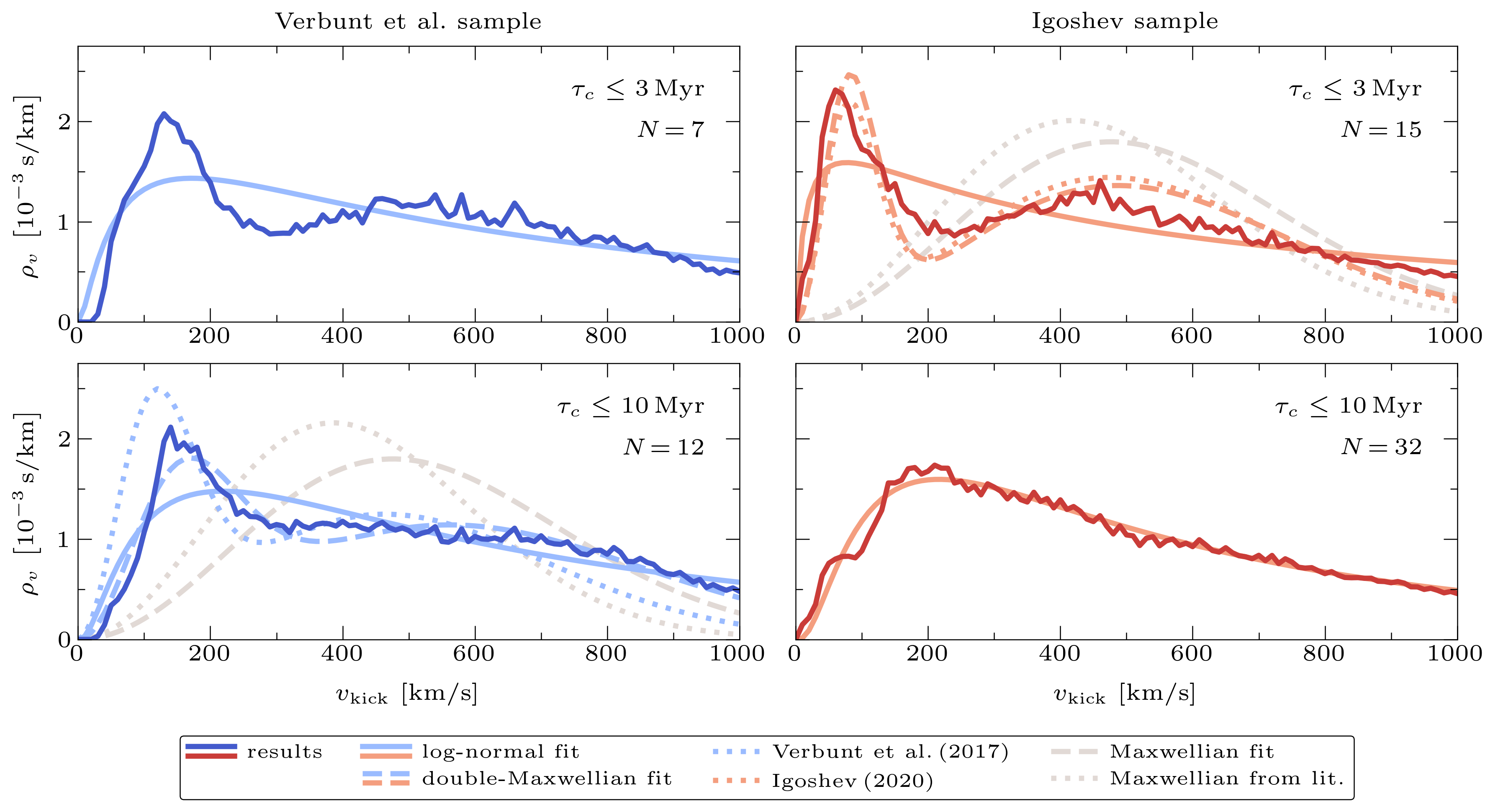}
    \caption{Resulting kick estimates (blue and red solid lines) for the young pulsars in the sample of \citet[][blue and left column]{Verbunt_2017} and the expanded sample of \citet[][red and right column]{Igoshev_2020}, similar to Fig.\ \ref{fig6} but excluded pulsars with $\tau_{\text{kin}}>\tau_{\text{c}}+5$ Myr. We fitted log-normal distributions (solid light blue/red) to the results in each panel, and to the results corresponding to the ages analysed by the literature we also fitted Maxwellians and double-Maxwellians (light blue/red and grey dashed lines). The corresponding distributions posed by \citet{Verbunt_2017} and \citet{Igoshev_2020} are also shown (light blue/red and grey dotted lines).}
    \label{figD}
\end{figure*}
\section{Kinematic age selection}
\label{App.D}
In order to investigate whether the pulsars with $\tau_{\text{kin}}>\tau_{\text{c}}+5$ Myr (as marked in Fig.\ \ref{fig2}) cause the bimodality in the young pulsar sample (as shown in Fig.\ \ref{fig6}), we determined the kick distributions for the young samples while excluding these pulsars. In Fig.\ \ref{figD} we show the kick distributions of young pulsars with aligning $\tau_{\text{c}}$ and $\tau_{\text{kin}}$. In these kick distributions the bimodality is still present, meaning it is unlikely caused by the pulsars with inconsistent age estimates.
\section{Igoshev sample}
\begin{figure}[h]
    \resizebox{\hsize}{!}{\includegraphics{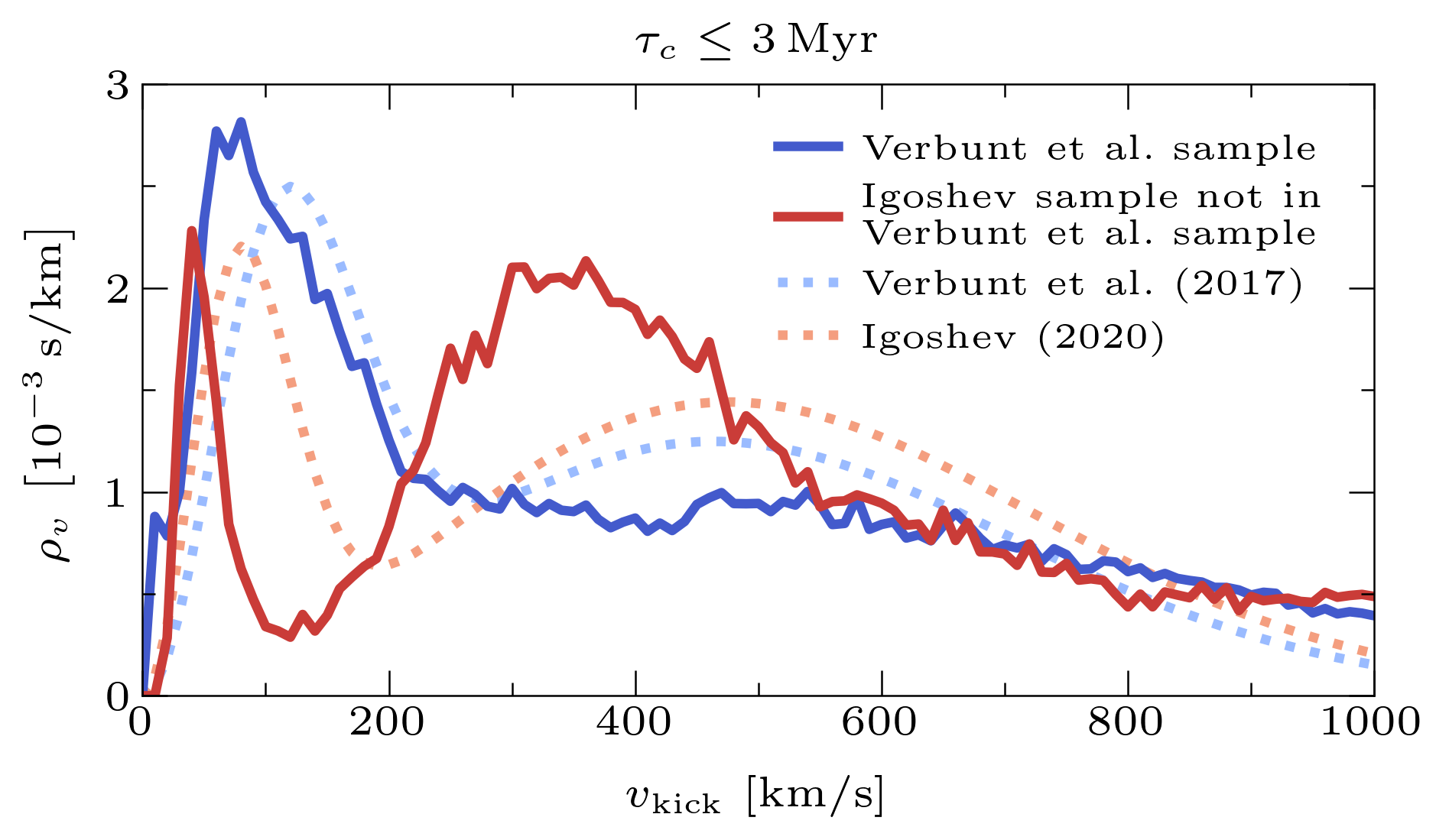}}
    \caption{Kinematically constrained kick distributions for the sample of \citet[][blue line]{Verbunt_2017} and the sample of pulsars that are only in the \citet[][red line]{Igoshev_2020} sample \citep[i.e.\ the observations of][]{Deller_2019}, for $\tau_{\text{c}}\leq3$ Myr. The dotted lines show the kick distributions posed by \citet{Verbunt_2017} and \citet{Igoshev_2020}, where we note that the former is based on pulsars with $\tau_{\text{c}}\leq10$ Myr instead of $\tau_{\text{c}}\leq3$ Myr.}
    \label{figE}
\end{figure}

The two pulsar samples used in the main analysis of this paper are the samples of \citet{Verbunt_2017} and \citet{Igoshev_2020}, where the Igoshev sample expands the Verbunt et al.\ sample with the results of \citet{Deller_2019}. The question arises whether the pulsars that were added to the Verbunt et al.\ sample to form the Igoshev sample follow a kick distribution similar to the results of \citet{Verbunt_2017}. In Fig.\ \ref{figE} we compare the kinematically constrained kick distributions of the Verbunt et al.\ sample to the kick distributions for the pulsars that are only present in the Igoshev sample. The resulting distributions do not describe a consistent kick distribution: the Verbunt et al.\ kicks peak at ${\sim}100$ km/s, whereas the expansion of the Igoshev sample has a dearth at ${\sim}100$ km/s and peaks at ${\sim}50$ km/s and ${\sim}300$--$400$ km/s. This could either be caused by low-number statistics or perhaps an observation bias in the observations of \citet{Deller_2019}.
\end{appendix}

\end{document}